\documentclass{article}
\usepackage{epsfig,calc,amssymb,amsfonts,natbib,url,xspace,listings,texdraw,thrmappendix}
\usepackage[utf8]{inputenc}

\usepackage{fancybox}
\usepackage{array}

\usepackage[x11names, rgb]{xcolor}
\usepackage{tikz}
\usetikzlibrary{arrows,shapes}
\usepackage{gnuplot-lua-tikz}
\include{gnuplot-lua-tikz}
\usepackage{amsmath}

\def\tighten{ }

\long\def\comment#1{}

\def\ttab{\hspace{.5cm}}

\def\definevector#1{\expandafter\gdef\csname #1v\endcsname{{\bf #1}\xspace}}

\definevector{A}\definevector{B}\definevector{C}\definevector{D}\definevector{E}\definevector{F}\definevector{G}\definevector{H}\definevector{I}\definevector{J}\definevector{K}\definevector{L}\definevector{M}\definevector{N}\definevector{O}\definevector{P}\definevector{Q}\definevector{R}\definevector{S}\definevector{T}\definevector{U}\definevector{V}\definevector{W}\definevector{X}\definevector{Y}\definevector{Z}\definevector{a}\definevector{b}\definevector{c}\definevector{d}\definevector{e}\definevector{f}\definevector{g}\definevector{h}\definevector{i}\definevector{j}\definevector{k}\definevector{l}\definevector{m}\definevector{n}\definevector{o}\definevector{p}\definevector{q}\definevector{r}\definevector{s}\definevector{t}\definevector{u}\definevector{v}\definevector{w}\definevector{x}\definevector{y}\definevector{z}

\def\definemathvector#1{\expandafter\gdef\csname #1v\endcsname{\mbox{{\boldmath$\csname #1\endcsname$}}}}

\definemathvector{alpha}  \definemathvector{beta}  \definemathvector{gamma}  \definemathvector{delta}  \definemathvector{epsilon}  \definemathvector{iota}  \definemathvector{mu}  \definemathvector{theta}  \definemathvector{sigma}  \definemathvector{Sigma}               

\long\def\longcodefig#1#2#3{
\setcounter{lstlisting}{\arabic{figure}}
\begin{figure}
\stepcounter{figure}
\lstset{showstringspaces=false, basicstyle=\footnotesize, emphstyle=\bfseries, language=C,%
breaklines=true,%
caption={#2},captionpos=b,label=#3}
\vskip 5pt
\hrule
\lstinputlisting{#1.c}
\hrule
    \vskip 0.5cm
    \setlistdefaults
\end{figure}
}
\def\ttab{\phantom{hell}}

\def\setlistdefaults{\lstset{ showstringspaces=false,%
 basicstyle=\small, language=C, breaklines=true,caption=,label=%
,xleftmargin=.34cm,%
,frameshape=
,frameshape={nnnynnnnn}{nyn}{nnn}{nnnynnnnn}
}
\lstset{columns=fullflexible, basicstyle=\small, emph={size_t,apop_data,apop_model,gsl_vector,gsl_matrix,gsl_rng,FILE,math_fn,an_agent,agent_s},emphstyle=\bfseries}
}
\setlistdefaults

\newenvironment{items}{
\setlength{\leftmargini}{0pt}
\begin{itemize}
  \setlength{\itemsep}{3pt}
  \setlength{\parskip}{0pt}
  \setlength{\parsep}{3pt}
}{\end{itemize}}

\usepackage{times, epsfig, latexsym, setspace}
\usepackage[T1]{fontenc}

\newcounter{apopctr}

\def\Con{{\rm Con}}
\newcommand{\RNG}{{\textsc{RNG}}\xspace}
\newcommand{\CDF}{{\textsc{CDF}}\xspace}
\newcommand{\PMF}{{\textsc{PMF}}\xspace}
\renewcommand{\L}{{\textsc{L}}\xspace}
\newcommand{\LL}{{\textsc{L}}\xspace}
\newcommand{\Est}{{\textsc{Est}}\xspace}

\def\Re{{\ensuremath{\mathbb R}}\xspace}
\def\Cat{{\ensuremath{\mathbb C}}\xspace}
\def\Nat{{\ensuremath{\mathbb N}}\xspace}

\def\Datapt{\ensuremath{\Dv}\xspace}
\def\datapt{\ensuremath{\dv}\xspace}
\def\parampt{\ensuremath{\pv}\xspace}
\def\models{\ensuremath{{\mathbb M}}\xspace}
\def\datas{\ensuremath{{\mathbb D}}\xspace}
\def\params{\ensuremath{{\mathbb P}}\xspace}

\def\paramsub{\ensuremath{\underline{\mathbb P}}\xspace}
\def\mod#1{\ensuremath{M_{#1}}\xspace}

\def\code#1{{\tt #1\xspace}}
\errorstopmode

\begin{document}

\author{Ben Klemens
\thanks{The author thanks Amber Baum,
Aref Dajani, Derrick Higgins, Guy Klemens, Joshua Tokle, Lynne Plettenberg, Andrew Raim, Rolando Rodr\'iguez, Anindya Roy, Joseph Schafer, and Tommy Wright for comments
and support. 
This paper first appeared in July 2014 as a working paper in a U.S.\ Census Bureau
research report series.
This report is released to inform
interested parties of ongoing research and to encourage
discussion. The views expressed are
those of the author and not necessarily those of the U.S.
Census Bureau.
}
}
\title{A useful algebraic system of statistical models}

\maketitle

\abstract{
This paper proposes a single form for statistical models that accommodates a broad range of models,
from ordinary least squares to agent-based microsimulations.  The definition makes it
almost trivial to define morphisms to transform and combine existing models to produce
new models. It offers a unified means of expressing and implementing methods that are
typically given disparate treatment in the literature, including transformations via
differentiable functions, Bayesian updating, multi-level and other types of composed
models, Markov chain Monte Carlo, and several other common procedures.  It especially
offers benefit to simulation-type models, because of the value in being able to build
complex models from simple parts, easily calculate robustness measures for simulation
statistics and, where appropriate, test hypotheses. Running examples will be given
using {\em Apophenia}, an open-source software library based on the model form and transformations
described here.

}

\section{Introduction}
This paper presents a formal definition of the informally-understood concept of
statistical models and discusses the benefits of that definition.

The problem is not that there is no such formal definition, but that there are too
many, as each subfield of statistics develops its own. The definition here
hopes to strike a balance between those definitions that are structured to fit only
one genre of modeling, and definitions that are so unstructured that they provide
little advantage to users.

This paper describes a model as a collection of elements including a data space, a
parameter space, and functions mapping over those spaces: estimation, likelihood,
random number generation (RNG), and cumulative distribution function (CDF).  

The formalization provides sufficient structure to allow the definition of
consistency rules among the components of a model and well-defined transformations of
models. Formally, the set of models combined with the transformations form an algebraic
system. Basic models and transformation routines are the nouns and verbs of a modeling
language allowing authors to express and analyze complex models.

Informally, researchers often describe a situation to be described via a narrative,
maybe something like: {\em first, a Normally-distributed random event occurs, then
it interacts in a specific way with another random event with a Beta distribution to
produce an aggregate, then the outcome is a linear transformation of the aggregate.} But
it can be difficult to formalize the narrative into something that can be estimated and
compared to other narratives or a data set. The algebraic system solves the problem,
by providing a set of mappings that each take in a well-formed model (or models),
and outputs a well-formed model. When each segment of the narrative is a model, it is
easy to build the full storyline by applying successive transformations.

In contrast, a simple transformation of the structures defined in many textbooks and
built in to common modeling platforms may produce something outside of the space of
models defined by the platform or textbook. Such inegalitarian treatment of models
causes a ``computational cliff'', where a slight variant of a built-in model (such as
replacing a Normal distribution with a truncated Normal) may require an entire rewrite
of procedures written around the built-in model. The computational cliff encourages
researchers to use slightly inappropriate stock models rather than modify them for
the situation at hand.

More broadly, authors may provide many tools for their specific favored genre of
modeling and relegate others to a can-be-implemented status.  For packages where
agent-based modeling (ABM) is a first-class paradigm, including Repast \citep{repast},
Sugarscape \citep{gass}, Netlogo \citep{netlogo}, and several Java libraries,
implementing an ordinary least squares (OLS) regression is technically possible but
practically infeasible.  Conversely, OLS is a first-class modeling paradigm in any
of a number of packages such as R, but it is technically possible but
practically infeasible to implement nontrivial ABMs in R.\footnote{The R community
provides support for this claim: the CRAN (Comprehensive R Archive Network, modeled on \TeX{}'s
CTAN and Perl's CPAN) has several thousand packages, but a search for agent modeling packages turned
up only packages like Rnetlogo, which use R as a front-end to systems where ABMs are
first-class models.}

By writing tools around a standard, generic form for models,
it becomes almost trivial to use tools across modeling traditions, such as applying
traditional statistical tools like maximum likelihood methods, variance estimation,
and appropriate hypothesis tests to agent-based or machine learning models.
This paper will include several examples that cross modeling traditions:

\begin{itemize}
         \setlength{\itemsep}{1pt}
        \setlength{\parskip}{0pt}
\item Example 4 shows how Ordinary Least Squares can be fit into the model structure here.

\item Example 5 presents a simulation of the production of a network of nodes, and
describes the model in the same model structure.

\item Example 8 compares the link distributions generated by the model of Example
5 to an Exponential distribution, and finds finds the best-fitting parameter for
the Exponential distribution.

\item Example 10 describes a mixture of populations from distinct distributions, and sets
Bayesian priors on the parameters describing the various populations.

\item Example 13 starts with a consumer preference model, wherein agents search for the optimal
bundle of goods given a utility function and a budget constraint. The model is
transformed via a Bivariate Normal kernel smoother, and the optimal parameters of
the transformed model are calculated given a fixed data set.

\item Example 14 is a model composed via an agent-based spatial search model as the
data-generation process and a Weibull likelihood function for evaluation.  By putting
priors on the search model parameters, we find a posterior distribution over the
Weibull parameters.
\end{itemize}

The definition of a model and the set of transformations built around it provide a
set of forms to be filled out for each model.  For standard closed-form models, the
blanks in the forms can be filled in with closed-form expressions. For other models,
each blank can be filled in using a well-defined computational algorithm. The process
is therefore egalitarian in the sense that all parts of every form are complete
for all models, but it remains inegalitarian in that the closed-form expression has
arbitrary precision and computational algorithms are always an approximation. For some
model--algorithm combinations, a great deal of computing time is required to achieve a
reasonable approximation; it is up to the model user to recognize these situations and
accommodate them with more hardware and time, model simplifications, or by contributing
a more efficient algorithm to the literature.

Are there theoretical constraints to treating an arbitrary objective function, as
given by a simulation or other mechanism, as a statistical model? For probability
distributions defined as such, the integral of the likelihood over the data space
given fixed parameters is defined to equal one (herein {\em the unitary axiom}). For
an arbitrary objective function, the integral over all data may take any value. However,
the unitary axiom is unnecessary for most of our purposes, where a simple ratio of
likelihoods or other relative comparison is all that is needed. Because calculating the
total density for a distribution so that it can be normalized to conform to the unitary
axiom is an expensive operation (it is the motivation for a great deal of Bayesian
computational machinery), we do so only when necessary.

Apophenia \citep{klemens:modeling} is an open-source library of routines that implements several
dozen models in a form similar to the one described here, and provides several functions
and other transformations that make use of such models. Apophenia demonstrates that the
concepts described in this paper can be put to practice: it was used to fit the running
examples in this paper, and, as a back-end to functions written for an R front-end,
is used by the U.S. Census Bureau for some disclosure avoidance operations on the
2010 Census and American Community Survey.  An appendix provides further discussion
of Apophenia and a set of full examples; the code used to generate the examples is
available at \url{http://github.com/b-k/modeling_examples}.

However, this paper has been written to be as platform-independent as possible,
and readers who are tied to an existing computing platform should find much in this
paper that could easily be implemented using the tools they have available. Readers
who expect that the model structure and transformations described herein are easy to
reimplement via their preferred platform are encouraged to do so.

Section \ref{priorworks} considers the history of
prior attempts to produce systems that cover a broad range of statistical models,
and even the problem of defining the word {\em model}.

Section \ref{modeldefsec} presents the model definition used by this paper, as a
collection of functions that either take in or output parameters or data.

Given a well-defined set of models, it is possible to define morphisms mapping from the
space of models back to the space of models. Section \ref{transforms}
presents several examples.

Section \ref{applicationsec} considers some immediate applications of a standard
model form, such as means of hypothesis testing for an arbitrary model.

The appendix provides several complete examples in code. They are not pseudocode, but
working examples, demonstrating how models and model transformations would be implemented
in practice, some egalitarian uses of slightly nonstandard models, a complex model
built by composing simple models via morphisms, and two agent-based microsimulations.

\section{Prior works}\label{priorworks}

In a 2007 talk about the future of statistics, Bradley Efron described the present
day as ``an era of ragtag heuristics, propelled with energy but with no guiding
direction.'' \citep{efron:futuretalk} Indeed, a modern researcher has many different
interpretations of the word {\em model} at his or her disposal, each backed by a research
community propelled with energy in its own direction: generalized linear models (GLMs),
simple distributions like the Normal or Dirichlet, hierarchical nests of models, Markov
chains, Bayesian frameworks, discrete event simulation, agent-based simulation, and so
on. But, in Efron's words, we are not yet seeing these ragtag heuristics ``coalescing
into a central vehicle for scientific discovery.''

\subsection{Computing literature}
Demand for a unified model framework seems to have stemmed mostly from designers of new
computing systems as they confronted the problem of building model objects.

Some designs take existing formal systems and add a probabilistic element,
culminating in stochastic programming languages such as Church \citep{church}, BLOG
\citep{blog}, and Venture \citep{venture}. Models focus on tracking states of the world, from which Bayesian nets
(akin to many structural equation modeling tools) can be formed. It would be easy
and natural to implement the algebraic system of models described in this paper
using these stochastic programming languages.

The HLearn package for the Haskell programming language \citep{hlearn} casts a variety
of statistical models in terms of underlying algebraic structures for the purposes
of parallelizing their implementation. It restricts itself to model transformations
expressible as monoids.  \citet{monoidify} also goes into detail about the value of
a strict algebraic structure, using simple statistics as examples.

BUGS (Bayesian Inference Using Gibbs Sampling, \citet{bugs}) has been reimplemented several times in
several variants [JAGS (Just Another Gibbs Sampler), OpenBugs \citep{openbugs}], each of
which has a common flavor. As per the names, the focus is on chaining together models
via Gibbs sampling. These packages therefore offer strong model-composition features,
but make little effort to provide an egalitarian model object.

Object-oriented systems like C++ or Java require a class or object declaration,
which defines the form to which all objects in the class must conform.  Zelig
\citep{imai:zelig} follows this lead, defining a class format into which it places a
large set of contributed models.  The class structure simplifies the workflow for models
that explain a single outcome variable using a set of input variables---basically,
the traditional generalized linear model (GLM) framework and related models.

Lisp-Stat \citep{tierney:lispstat} follows the mold of object-oriented
systems based on an inheritance structure, where specific models inherit from a
relatively abstract model. Specific types of GLM, for example, would inherit from
a GLM prototype (which \citet{tierney:glm} implements). A model may respond to any
from a long list of messages, including simple requests like \code{:coef-estimates}
to get the estimated coefficients, up to more specialized and computation-intensive
commands like \code{:cooks-distance}.

The lead author of Lisp-stat explicitly rejects class formats as too restrictive
\citep[p 206]{tierney:lispstat}. Instead, a model is a Lisp-style list of data or
procedures, and authors may add as many new items to the list as are useful, without
class-defined constraints.  Thus, the system's key strength is that it is flexible,
and new functions may easily be inserted into any model structure.

For our purposes, the system's key failing is that it is flexible, and new functions may
easily be inserted into any model structure. Each class of model will have its own set
of functions that make sense for its genre of modeling, so there is no guarantee that
the internals of any two models will match sufficiently well that one could be swapped
for another in a given procedure. If we wish to apply a transformation to a
model, we will need to modify every interface function in the model, which we can do
only if we have a comprehensive and consistent list of such functions.

The problem, then, is to define a class structure that is broad enough
to usefully describe any model, but sufficiently limited that we can
be guaranteed that every model implements every element included in
the definition.

\subsection{Theory literature}
Away from the computer, there is largely consensus on the definition of a model.

Begin with a data space, \datas, and a parameter space, \params. 
In practice, these spaces are typically a space of $N$-dimensional real numbers, $\Re^N$;
a sequence of discrete categories, $\Cat_1\times \dots \times \Cat_N$ (e.g., sex $\times$ age bracket $\times$ race $\times$\dots); a $P$-dimensional sequence of
natural numbers $\Nat^P$; or a Cartesian product of these spaces.
The numeric values typically have units
attached. The parameters may be a list of not-predetermined length; if each element
of the list is in $\Re$ then the full list is in the space consisting of the set of sets
$\{\emptyset, \Re, \Re^1, \Re^2, \dots\}$; a model over a parameter space whose elements
do not have predetermined dimension is referred to using the counterintuitive but
customary term {\em non-parametric model}.  In some cases, one or both of $\datas$
and $\params$ may be the empty set.  Elements of a given data space $\datas_m$ will
be written as $\datapt_m$; elements of a given parameter space $\params_m$ will be
written as $\parampt_m$.

The $\datas$ space will need to be totally ordered (that is, a $\leq$ operation is
defined so that $\datapt_1\leq \datapt_2$ or $\datapt_2\leq \datapt_1, \forall \datapt_1,
\datapt_2\in \datas$), for the purposes of the cumulative density function (CDF), which
is primarily for hypothesis testing. Even for cases like unordered categories, the CDF
can still have real-world meaning. For example, given unordered categories $A, B, C, D$
and imposed alphabetical ordering, $\CDF(C) - \CDF(B)$ is the density on only category $C$.

\citet{mccullagh:what} gives a long list of authors who either explicitly
or implicitly use a definition of parameterized statistical model akin to the
following, given a data space $\datas$, a space of parameter sets $\params$, and the
nonnegative real numbers $\Re^+_0$:

\stmt{defn}{modeldefone}{

A parameterized statistical model is a parameter set
$\params$ together with a function $\params \to (\L:\datas\to\Re^+_0)$, which
assigns to each parameter point $\parampt \in \params$ a probability
distribution $\L_{\parampt}$ on $\datas$.
}

See also \citet[p 119]{hill:gut}, who gives another definition following this mold.

A Normal model in this context is a function ${\cal N}:(\mu,\sigma)\to (\L:\Re\to\Re_0^+)$, where $\mu$ is read as the mean, and $\sigma$ the standard deviation. One
could also express the mapping via a single function in three
variables, $\L(\datapt, \mu, \sigma)$, where $\datapt$ is a scalar data point.
Conditioning on a single point in the parameter
space, such as $(\mu=0$, $\sigma=1$), fixes a single likelihood function, $\L(\datapt,
\mu, \sigma| \mu=0, \sigma=1)$, which is a function of only $\datapt$.\footnote{ Given a
function $\L(\cdot, \cdot)$ taking in data $\datapt$ and parameter vector $\parampt$ and reporting the likelihood of the pair, one could fix $\parampt$ and thus generate what
is called a probability function, $L(\cdot, \parampt)$. Or, one could fix $\datapt$
and produce what is called a likelihood function, $\L(\datapt, \cdot)$. RA Fisher
explains that the two should remain separate in interpretation:

\begin{quote}
\dots [I]n 1922, I proposed the term `likelihood,' in view of the fact that, with
respect to [the parameter], it is not a probability, and does not obey the laws of
probability, while at the same time it bears to the problem of rational choice among
the possible values of [the parameter] a relation similar to that which probability
bears to the problem of predicting events in games of chance\dots. Whereas, however,
in relation to psychological judgment, likelihood has some resemblance to probability,
the two concepts are wholly distinct\dots.''

\hfill         ---\citet[p 287]{fisher:likelihood}
\end{quote}

However, setting a rule that $\L(\cdot, \parampt)$ should be interpreted differently from
$\L(\datapt, \cdot)$ does not help us determine what to call $\L(\cdot, \cdot)$ itself.
Further, the computer is entirely indifferent to the distinction between the ostensibly
objective and subjective views of $\L(\cdot, \cdot)$. In this paper, I use the notation
$\L$ instead of \textsc{P} to reduce ambiguity with the parameter space, for all variants
of $\L(\cdot, \cdot)$.}

Definition \ref{modeldefone} casts a model as a simple collection of spaces and mappings
between those spaces, consisting of $$\left(\datas, \params, \Re^+_0,
        \L:(\datas, \params)\to\Re^+_0 \right).$$

We have a function that takes inputs from both $\datas$ and $\params$, but
one could add other functions to the collection including the sets $\datas$, $\params,$
and $\Re^+_0$, including functions that take inputs only from $\datas$ and output to $\params$,
such as a routine to estimate the most likely model parameters; or functions that
take in only an element of $\params$ and output to \datas, such as an expected value
calculation or a random number generator.

This paper argues that there is real benefit to including some of these other functions
in the collection defining a model. 

\section{Definitions and examples} \label{modeldefsec}
Here is the definition of a model that will be used for the remainder of the paper:

\stmt{defn}{modeldef}{
A {\em model} is a collection consisting of the sets $\datas$, $\params$, $\Nat$, and $\Re^+_0$ and the
following mappings between sets:

\begin{itemize}\tighten
\item Likelihood: $\L:(\datas, \params) \to \Re^+_0$. 
\item Estimation: $\Est:\datas \to \params$.
\item (Pseudo-)random number generator: $\RNG: (\params, \Nat) \to
\datas$.
\item Cumulative distribution function: $\CDF:(\datas, \params) \to [0,1]$.

\end{itemize}
}

The $\L$ function might also be described as an {\em objective function}, which
has fewer associations than the term {\em likelihood}.

A (pseudo-)random number generator (RNG or PRNG) is a determinstic function of input
parameters and an input seed, typically a large natural number. The RNG function
then deterministically maps each $(\parampt\in\params, n\in \Nat)$ combination to a
relatively unpredictable element of $\datas$.  With no second argument, $\RNG(\parampt)$
indicates a representative draw $\datapt\in\datas$.

Some definitions below will use $\int_{\delta\in\datas} L(\delta, \parampt) d\delta$,
which requires that $\datas$ have a metric such that the integral over the entire
space is well-defined. If $\datas$ is discrete, it will be understood that
$\int_{\delta\in\datas}$ is the sum $\Sigma_{\delta\in\datas}$. Definition \ref{LofD} will
require that $\int_{\rho\in\params} L(\datapt, \rho) d\rho$ (or the corresponding sum) be
well-defined.

A {\em data set}, written as $\Datapt$, is a collection of $N$ elements from a single
data space, where $N$ is any integer $\geq 1$. If the $N$ elements
are independent, then for any likelihood function $p(\cdot)$, we define $p(\Datapt)
= p(\datapt_1)p(\datapt_2)\cdots p(\datapt_N)$, so the sequel will primarily discuss
likelihood functions over a single data element.

This model definition is a mix of mathematics and the practical reality of how models
are used, and later sections will demonstrate why a `larger' definition of a model has
practical benefit. For example, the CDF is included because hypothesis tests are
typically an assertion regarding some portion of the CDF.  Even so, a real-world
implementation as a computational structure or object would likely digress from
this bundle of functions. One might prefer to include a log likelihood
function, rather than the plain likelihood function listed above, or might include
an entropy function, even though entropy can be calculated using the elements already
given.

We are interested only in models that have basic internal consistency.

\stmt{defn}{consistentdef}{ A model is {\em ML-consistent} iff:

\begin{itemize}\tighten
\item $\int_\datas L(\delta, \parampt) d\delta$ is greater than zero and finite for any fixed $\parampt\in\params$ iff $\datas \neq \emptyset$.
\item For given datum $\datapt$, $argmax_p L(\datapt, p) = \Est(\datapt)$. That is, the estimation
function of a model produces a maximum likelihood estimate (MLE) of the model's likelihood
function.
\item The odds of drawing a data point from the RNG is proportional to its likelihood.
That is,
for any fixed $\parampt\in\params$ and any $\datapt_1, \datapt_2 \in \datas$, 
$($probability of drawing $\datapt_1=\RNG(\parampt))\cdot L(\datapt_2, \parampt)
=
($probability of drawing $\datapt_2=\RNG(\parampt))\cdot L(\datapt_1, \parampt)
$. 

\item For any fixed $(\parampt\in\params, \datapt\in\datas)$, the probability that $\RNG(\parampt) \leq \datapt$ equals $\CDF(\datapt, \parampt)$.

\end{itemize}
}

Let \models indicate the space of ML-consistent collections. This is the set of models
that this paper considers.  Elements of \models will be distinguished by a subscript, such
as the Normal distribution model, \mod{\cal N}, and where needed the constituent functions will be given
matching subscripts: $\L_{\cal N}$, $\Est_{\cal N}$, $\RNG_{\cal N}$, and $\CDF_{\cal N}$.

There is typically an additional requirement (one of the Kolmogorov axioms) that
$\int_\datas L(\delta, \parampt) d\delta =1$ for any fixed $\parampt$; note that such
a constraint is not imposed here.  However, because the \CDF function conforms to
the \RNG function, which is proportional to the \L function, CDF(\datapt, \parampt) approaches one
as $\datapt$ gets larger, for any $\parampt$. Dropping the unitary axiom simplifies most of the analysis,
and Section \ref{l_of_p} will consider options for those cases where we need to know
$\int_\datas L(\delta, \parampt)d\delta$.

The consistency conditions for the CDF and RNG are largely definitions of these
terms. The condition that $\Est(\datapt)$ is an MLE of the likelihood
function given the data is nontrivial. Maximum likelihood has desirable properties,
like how it is an unbiased estimator as the data size $\to \infty$ under weak and
common assumptions, and has undesirable properties, like how it is often a biased
estimator for small samples \citep{yudi:likelihood}.

One can easily find useful models in the literature where the estimation routine does
not produce an MLE, such as estimations based on minimizing mean-squared
error (MSE). One could define {\em MSE-consistency} as similar to Definition
\ref{consistentdef} but with the MLE condition replaced with estimation via minimized
MSE. This would define a new space $\models_{mse}$, with a new set of transformations
mapping from $\models_{mse} \to \models_{mse}$.

Other potentially interesting rules could include Kullback-Leibler divergence
minimizing consistency, entropy-maximizing consistency, or a method of moments-style
rule matching model moments to data moments. Implementing a system of modeling based on
KL-consistency, entropy-consistency, or any other option is left as an exercise for the reader.

\paragraph{Example 1: The Normal distribution}

Closed-form or computationally-efficient methods have been derived for every function in
the model definition:

\begin{items}
\item $\datas= \Re$.
\item $\params=\Re\times \Re^+$, representing the mean $\mu$ and standard deviation $\sigma$.
\item Likelihood: ${\cal N}(\mu,\sigma) = {\frac{1}{\sqrt{2 \pi \sigma^2}}} \exp
(\frac{-(x-\mu)^2}{ 2\sigma^2})$. 
\item Estimation: $\hat\mu =$ mean of $\Datapt$; $\hat\sigma= \sum_{\datapt\in\Datapt}(\datapt - \hat\mu)^2/n$.
\item RNG: No closed form expression, but see \citet{devroye:rng}.
\item CDF: No closed-form expression, but see \citet{recipesinc}.
\end{items}

Let \mod{\cal N} indicate this model.

\paragraph{Example 2: The Probability Mass Function (PMF)}

 A data set can be treated as a probability
distribution, where the likelihood of a given observation is proportional to the
frequency with which the observation appears in the data. The PMF model will appear
frequently throughout this paper, when a closed-form model can not be calculated or derived.

Let $\Datapt_I$ be the initial data set used to estimate the PMF, and its individual
elements $\datapt_{I1}, \datapt_{I2}, \dots, \datapt_{IN}$.

\begin{items}
\item \datas: Typically $\Re^N$ or a short list of categories.
\item $\params=\Re^N\times \datas^N$, representing weights $(w_1, \dots, w_N)\in \Re^N$ assigned to $N$ observations from \datas.
\item Estimation: $\parampt\equiv \left((w_1, \datapt_{I1}), (w_2, \datapt_{I2}), \dots, (w_N, \datapt_{IN})\right)$.
\item $\L(\datapt, \parampt)$: If $\datapt\not\in \Datapt_I$, zero.
Else, the  weight assigned to $\datapt$.
\item RNG: weighted draw from $\parampt$.
\item CDF(\datapt, \parampt): sort $\datapt_I$s according to $\leq$; sum weights up to $\datapt$.

\end{items}

Because of the format of \params, a model based on four observations in $\Re$ will
be different from one based on five observations in $\Re$, so this template defines
a multitude of elements of \models.  This paper will use $\mod{PMF}$ to refer to any
of these models.

In some cases, kernel smoothing or moving average transformations can improve how
well a PMF represents the underlying process.

\paragraph{Example 3: Coin flip}

A coin-flip model where heads and tails are equiprobable could be expressed via a
PMF estimated using the two-observation data set $\Datapt_I = \{{\rm heads}, {\rm
tails}\}$. Then $\parampt=\Est(\Datapt_I) = \{(\frac{1}{2}, {\rm heads}), (\frac{1}{2}, {\rm tails})\}$.

\paragraph{Example 4: Ordinary Least Squares}

 OLS is typically described via an equation $\Yv=\betav\Xv
+ \epsilon$, relating one subset of the input data, the `independent variables' $\Xv$,
to an output variable, the `dependent' $\Yv$; and $\epsilon$ is a Normally-distributed
error term. There are two considerations required to fit OLS into \models.

First,
the process of modeling for an OLS enthusiast consists of selecting a set of
independent variables.  As with \mod{PMF}, slightly different data spaces produce
different models all sharing a common template: an OLS model where \datas=[outcome,
age, sex, log(income)] is distinct from one where \datas=[outcome, age,
log(income)]. The model is a joint distribution across the full data
space including both independent and dependent variables.

Second, the likelihood of a given point $(\Yv, \Xv)$ under the OLS model is
calculated using the likelihood from the Normal model, $\L_{\cal N}\left((\Yv -
\betav\Xv), \sigma\right)$ \citep[p 144]{greene}. The integral of this likelihood over all
possible values of $(\Yv, \Xv)$ is infinite, so the development here adds additional
structure to reduce the likelihood to a finite integral, using 
a PMF model calculated using the $\Xv$ portion of the input data set. There are
infinitely many other developments possible.

\begin{items}
\item \datas: as specified by the model author, with dimension $N$.
\item $\params= \Re^N\times \Re^+_0$, indicating $(\betav, \sigma)$, including a
    coefficient on a constant column {\bf 1} inserted as the first column of $\Xv$.
\item Estimation: $\betav = (\Xv'\Xv)^{-1}(\Xv'\Yv)$
\item Likelihood: $\left\{
\begin{matrix}
\L_{\cal N}\left((\Yv - \beta\Xv), \sigma\right) & \Xv\in \Datapt_X \cr
        0 & \Xv \not\in\Datapt_X
\end{matrix}\right.$
\item RNG: Set $\parampt_{\PMF}=\Est_{\PMF}(\Datapt_X)$; draw $\Xv=\RNG_{\PMF}(\parampt_{\PMF})$; draw $\epsilon$ from $\RNG_{\cal N}(0, \sigma)$; set
$\Yv= \betav\Xv+\epsilon$.
\item CDF: Write the distribution of $(\Yv - \beta\Xv)$  as the sum of 
independent Normals,
$V_0 \sim {\cal N}(Y, \sigma/N)$ plus
$V_1 \sim {\cal N}(-\beta_1 X_1, \sigma/N)$ \dots plus
$V_{N} \sim {\cal N}(-\beta_n X_n, \sigma/N)$; calculate total CDF accordingly.

\end{items}

\subsection{Filling in missing elements}
Consider two authors who wish to fit their models into the model form. The first
is working with an Exponential distribution, which has well-known methods for calculating
parameter estimates, the CDF, and so on. When implementing the model, this author will
want to make use of these efficient routines.

The second author has developed a new function, say a simulation or an eccentric
relationship between dependent and independent variables, has embodied it in a likelihood
function, and now wishes to ask questions that require estimating the most likely parameters
give a data set, or making random draws from the PDF implied by the likelihood function.
That is, the modeler wishes to use a full model but only has time to write down the
likelihood. This author will need a model implementation that fills itself in using
only one or two author-defined elements.

Every item from this point to the end of the paper will have a default routine that can be evaluated for any $M\in \models$ and some special-case routines for models where efficient shortcuts or closed-form solutions
exist. Because the purpose of this paper is to show that any type of model can be given
egalitarian treatment, the focus will be on the default algorithm, but in no case are
we obligated to use the default when there is a closed-form solution
available.

The `larger' collection of Definition \ref{modeldef} more readily handles sets of
default and model-specific cases than the `smaller' Definition \ref{modeldefone}. There
are hooks for model-specific estimation, RNG, and CDF routines, and defaults can be
provided in several directions:

\paragraph{$\L \to \Est$} Maximum likelihood estimation routines are
black-box routines that use the likelihood function regardless of its form
\citep{recipesinc}. These routines propose a parameter $\parampt_1$ and use the
likelihood function to calculate $L(\datapt, \parampt_1)$, then jump to further
samples $\parampt_2 \dots \parampt_n$ based on the values $\L(\datapt, \parampt_2),
\dots,\L(\datapt, \parampt_n) $. Thus, one can use these optimizers to implement an
estimation routine given only a likelihood function.  Search routines are designed
for the optimization of general functions, so probability-like properties that the
function being optimized integrate to one---or even that it evaluate to a nonnegative
value---are not required.

\paragraph{$\L\to \RNG$} 
In one dimension, adaptive rejection Markov sampling \citep{gilks:arms} uses likelihoods
as a black box to make draws from the likelihood at the correct rate. ARMS also
considers only the ratio of likelihoods, and therefore has no requirement that its
input function integrate to one. In multiple dimensions, Section \ref{mcmcrng} uses
a Metropolis-Hasting MCMC algorithm to draw from an arbitrary data space.

\paragraph{$\RNG \to \L$} Make a few thousand random draws from the 
RNG; write each down as a single observation in a new data set (this is the stochastic
variant of {\em memoization} of a function). Use the draws to estimate a PMF model
approximating the original model. The likelihood represented by the PMF is a rough discrete
approximation of the true likelihood, but it is consistent: as the
number of draws approaches infinity, the error goes to zero.
If \datas is continuous,
a smoothing transformation (cubic splines, kernel density, moving average) may improve the accuracy of the PMF.

The resultant PMF model has $\L$ and $\CDF$ methods, which may be used as approximations to
for the main model's $\L$ and $\CDF$.

\paragraph{$\RNG \to \CDF$}
Make random draws and count the percentage of draws less than the given point $\datapt$.

\paragraph{$\CDF \to \L$} Calculate \L via numeric deltas from \CDF.

\paragraph{$\CDF \to \RNG$} Given $\datas=\Re$, draw $r=$a random value from a
Uniform$(0, 1)$ distribution. Via binary search, secant method, or Newton-Raphson method,
locate $\datapt$ such that $\CDF(\datapt, \parampt)=r$.
Then $\datapt$ is an appropriately-weighed random draw from the model with \CDF
 \citep[pp 32--33]{devroye:rng}.

\vspace{.4cm}

\begin{figure}
\begin{center}
\scalebox{.6}{
\includegraphics{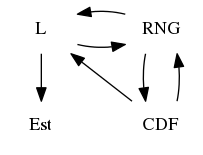}
}
\end{center}
\caption{A diagram of the listed means of filling in one element of a model with another.}\label{fillinflow}
\end{figure}

For most of these directions ($\L\to\Est$, $\RNG\to\L$, $\RNG\to\CDF$), the ML-consistency
requirements are met almost trivially; for the $\L\to\RNG$ and $\CDF\to\RNG$ algorithms,
see the given citations for discussion of how the algorithms generate an RNG that is
ML-consistent with the input likelihood and CDF.

Fiture \ref{fillinflow} shows the directions presented here, clarifying that one can begin
with any of \L, \RNG, or \CDF and arrive at any of the four function elements of the model
collection. The \Est element is lossy, in the sense that  one \Est function ignores a
great deal about the \L function away from the optimum, and so may
be equivalent to a variety of \L functions.

\paragraph{Example 5: A network-generation simulation}

This network generation simulation is based on agents with a position in an underlying
preference space (in this case, $\Re$). Each agent's preferences are first
generated as a draw from a ${\cal N}(0, 1)$ distribution (though $\sigma$ will
be allowed to vary in a variant below). Given two agents at positions $p_i$ and $p_j$,
they link with probability $1/(1+|p_i - p_j|)$. The output is a list of how many links
agents $1, \dots, N$ each have. It is useful to report sorted output, so the first
dimension is the highest link count, down to the final dimension reporting the lowest
link count.
See Figure \ref{randomnetmodel} for a summary of the algorithm.

\begin{figure}

\hrule
\begin{enumerate}
\item Fix $\sigma=1$.
\item For each agent $i$:
    \begin{enumerate}
    \item Draw a position $p_i$ using $\RNG_{\cal N}(0, \sigma)$.
    \end{enumerate}
\item For each pair of agents, $i$ and $j$:
    \begin{enumerate}
    \item Draw a random value $r \sim {\cal U}[0, 1]$.
    \item Link iff $r \leq \frac{1}{1+ |p_i-p_j|}$.
    \end{enumerate}
\item Output data $\leftarrow$ sorted count of connections for each person.
\end{enumerate}
\hrule

\caption{A simulation to produce a random distribution of link densities.}\label{randomnetmodel}
\end{figure}

\begin{figure}
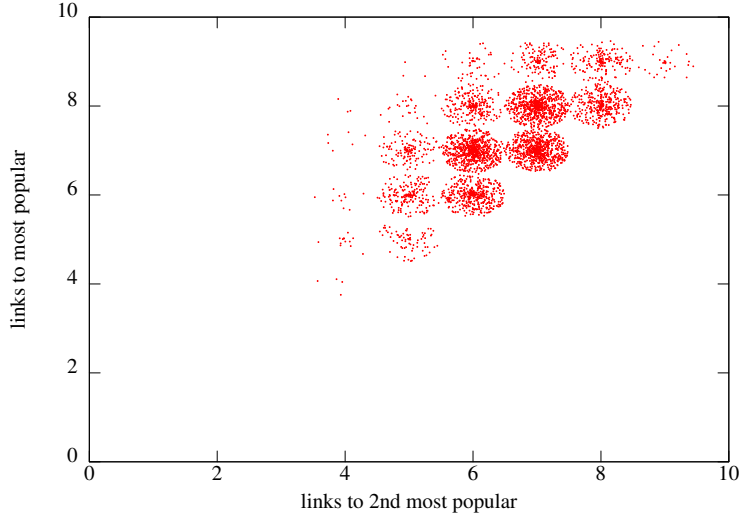

    \begin{center}
\scalebox{.8}{
\input just_sim.tex
}
    \end{center}
\caption{A distribution of the number of links to the two highest-ranked members of a ten-person network. Plot produced by running the simulation in Figure \ref{randomnetmodel}, then adding a jitter to each point.}\label{justsimplot}
\end{figure}

With $N=10$, one run of the simulation produces output in $\Nat^{10}$. Figure \ref{justsimplot} plots
the first two dimensions of the result of of 10,000 runs. For visualization purposes, a small jitter was added to
each point. 

The algorithm defines a model on $\params=\emptyset$ and $\datas=\Nat^{10}$.
We have only defined the RNG portion of the model, but the above transformations indicate
that the likelihood and other elements of the model can be well-defined by using the RNG.

Consider the hypothesis that the number of links to the most popular person is less
than or equal to four. The number of links for the second, third, \dots persons are
unrestricted, which we could express as the vacuous restriction that these link counts
are less than or equal to nine. One would evaluate this hypothesis using the percentage
of the total distribution that is below the point $[4, 9, 9,\dots, 9]$, which is
$$CDF_{\rm sim}([4, 9, 9,\dots, 9], \emptyset)\approx 0.0533.$$

Examples 13 and 14 in the appendix present some uses of more elaborate simulation models.

\section{Operations on models}\label{transforms}

This section presents a series of morphisms with signatures $f:\models
\to \models$ or $f:(\models, \models) \to \models$, that apply a well-defined
transformation to every element of the input model(s) to produce a new model, including
transformations via differentiable functions, fixing a parameter, or forming the product of 
independent distributions.

The transformations map to \models, so successive transformations can be chained together.
As the examples below will demonstrate, the transformation paradigm allows models of arbitrary
complexity to be built using simple steps.

\subsection{Conditional results}\label{l_of_p}
But before writing down these transformations, a few things can be said about when two
related models have equivalent elements, which will be
useful in checking that our transformations are ML-consistent and related the way their
descriptions claim they are.

\paragraph{The likelihood of \parampt}
These statements will rely on an additional set of definitions:

\stmt{defn}{LofP}{
    $$L(\parampt) \equiv \int_{\forall \delta \in \datas} L(\delta,\parampt)d\delta$$
and
$$L(\datapt|\parampt)\equiv L(\datapt, \parampt)/L(\parampt).$$
}

The unitary axiom requires that $L(\parampt) \equiv 1, \forall \parampt$; dropping
that assumption motivates the need for this definition.  Recall that the definition
of ML-consistency requires that $\infty > L(\parampt) > 0$ for all
$\parampt\in\params$.

\stmt{defn}{LofD}{
    $$L(\datapt) \equiv \int_{\forall \rho \in \params} L(\datapt,\rho)d\rho$$
and if $\infty > L(\datapt) > 0$, 
$$L(\parampt|\datapt) \equiv L(\datapt, \parampt)/L(\datapt).$$
}

This definition requires integrability on $\params$.

\paragraph{Equivalent elements} These definitions allow us to state when two models share
\Est, \RNG, or \CDF elements.

\stmt{lma}{preestinvar}{
If for each $\datapt\in\datas$ there exists a constant $K_\datapt$ such that
$${\L_2(\datapt, \parampt)} = K_\datapt {\L_1(\datapt, \parampt)}, \forall \parampt\in\params,$$
then $\Est_2(\datapt) = \Est_1(\datapt)$.
}

This is sufficient but not necessary.

\stmtproof{preestinvar}{
For an arbitrary $\datapt\in\datas$, $\parampt_{max}\equiv \Est_1(\datapt)$ and any other $\parampt_x$, 
\begin{eqnarray*}
    L_1(\datapt, \parampt_{max}) &\geq& L_1(\datapt, \parampt_x)\hbox{, so }\\
    K_\datapt\cdot L_1(\datapt, \parampt_{max}) &\geq& K_\datapt\cdot L_1(\datapt, \parampt_x)\hbox{, so }\\
L_2(\datapt, \parampt_{max}) &\geq& L_2(\datapt, \parampt_x).
\end{eqnarray*}
Thus, $\parampt_{max}$ is also
the argmax for $L_2(\datapt, \cdot)$, so $\Est_1(\datapt) =\Est_2(\datapt)$. The premise of
the lemma states that this holds for all $\datapt\in\datas$.
}

\vskip .2cm 
\pf \preestinvarproof \endpf
\vskip .2cm 

\stmt{prop}{estinvariance}{
If for each $\datapt\in\datas$ there exists a constant $K_\datapt$ such that
$${\L_1(\parampt|\datapt)} = K_d{\L_2(\parampt|\datapt)}, \forall \parampt\in\params,$$
then $\Est_1(\datapt) = \Est_2(\datapt)$.
}

\stmtproof{estinvariance}{
Because $L_1(\datapt)/L_2(\datapt)$ is constant for any given $\datapt$,
$\frac{L_1(\datapt, \parampt)}{L_2(\datapt, \parampt)}$ is a constant iff 
$\frac{L_1(\parampt|\datapt)}{L_2(\parampt|\datapt)}$ is a constant; then
\refstmt{preestinvar} applies.
}

\vskip .2cm 
\pf \estinvarianceproof \endpf
\vskip .2cm

\stmt{lma}{prernginvar}{
For each $\parampt\in\params$ there exists a constant $K_\parampt$ such that
$${\L_1(\datapt, \parampt)} = K_\parampt{\L_2(\datapt, \parampt)}, \forall \datapt\in\datas$$
iff  $\RNG_1(\parampt) = \RNG_2(\parampt)$ and $\CDF_1(\parampt) = \CDF_2(\parampt)$.
}

\stmtproof{prernginvar}{
Given that $L_1(\datapt, \parampt) = K_\parampt L_2(\datapt, \parampt)$.
By the definition of \RNG(\parampt), the ratio of the probability of drawing $\datapt_1$
to the probability of drawing $\datapt_2$ equals
$\frac{L_1(\datapt_1, \parampt)}{L_1(\datapt_2, \parampt)},$
and this ratio does not change when we multiply the numerator and denominator by
$K_\parampt$. The \CDF element of the model is defined in terms of the \RNG element.

In the other direction, $\RNG_1(\parampt) = \RNG_2(\parampt)$ for some $\parampt$ implies
that, for any arbitrary $\datapt_1$ and $\datapt_2$,

$$\frac{\L_1(\datapt_1, \parampt)}{\L_1(\datapt_2, \parampt)}
=
\frac{\L_2(\datapt_1, \parampt)}{\L_2(\datapt_2, \parampt)},$$
so 
$$\frac{\L_1(\datapt_1, \parampt)}{\L_2(\datapt_1, \parampt)}
=
\frac{\L_1(\datapt_2, \parampt)}{\L_2(\datapt_2, \parampt)}\equiv K_\parampt.$$}

\vskip .2cm 
\pf \prernginvarproof \endpf
\vskip .2cm

\stmt{prop}{rnginvariance}{
For each $\parampt\in\params$ there exists a constant $K_\parampt$ such that
$${\L_1(\datapt|\parampt)} = K_\parampt {\L_2(\datapt|\parampt)}, \forall \datapt\in\datas$$
iff $\RNG_1(\parampt) = \RNG_2(\parampt)$ and $\CDF_1(\parampt) = \CDF_2(\parampt)$.
}

\stmtproof{rnginvariance}{
$L_1(\parampt)/L_2(\parampt)$ is constant for a given $\parampt$. Then 
$\frac{\L_1(\datapt|\parampt)}{\L_2(\datapt|\parampt)}$ is constant iff 
$\frac{\L_1(\datapt, \parampt)}{\L_2(\datapt, \parampt)}$ is constant; then
\refstmt{prernginvar} applies.
}

\pf \rnginvarianceproof \endpf

The two propositions have a nice symmetry: proportional shifts in $\L(\parampt|\datapt)$ do
not affect $\Est$; proportional shifts in $\L(\datapt|\parampt)$ (which may differ from
$\L(\datapt, \parampt)$ because we dropped the unitary axiom) do not affect \RNG or \CDF.

\subsection{Parameter fixing} The simplest means of reducing an $N$-parameter model to an
$N-1$-parameter model is to fix a parameter at a constant value, such as turning a
two-parameter Normal distribution, ${\cal N}(\mu, \sigma)$, into a one-parameter
distribution, ${\cal N}(\mu, 1)$. One may also write this as ${\cal N}(\mu|\sigma=1)$.

\newcommand{\Fix}{{\textsc{Fix}}\xspace}
\begin{figure}
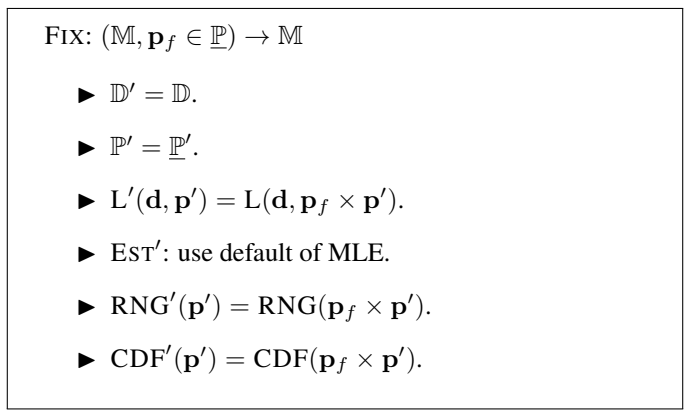

\begin{center}
\framebox[1.1\width]{
\begin{minipage}{8cm}
\vskip .1cm
\raggedright
\Fix: $ (\models,\parampt_f \in  \paramsub)\to \models $

\begin{itemize}\tighten
\def\labelitemi{$\blacktriangleright$}

\item $\datas'=\datas$.
\item $\params' = \paramsub'$.
\item $\L'(\datapt, \parampt')=\L(\datapt, \parampt_f\times \parampt')$.
\item $\Est'$: use default of MLE.
\item $\RNG'(\parampt') = \RNG(\parampt_f\times\parampt')$.
\item $\CDF'(\parampt') = \CDF(\parampt_f\times\parampt')$.

\end{itemize}
\vskip .1cm
\end{minipage}
}
\end{center}

\caption{Definition of the \Fix mapping.}\label{Fixfig}
\end{figure}

Figure \ref{Fixfig} shows the mapping from input model and constraint to
constrained output model. The mapping $\models\to\models$ breaks down into a mapping
of the separate components of a model: parameter space to parameter space, likelihood
function to likelihood function, and so on. The elements of the post-transform model
are given a prime, such as $\L'(\cdot, \cdot)$, $\Est'(\cdot)$, \dots.

Some of these transformations require additional information to be fully described,
which will be specified using subscripts. For example, a Normal model with $\sigma$
fixed at one will be written as $\Fix_{\sigma=1}(\mod{\cal N})$. These additional
details are ``private'' to the generated model, used for the inner workings of the
associated functions, but not used outside of the confines of the model.

Let the parameter subset be
fixed at $\parampt_f \in \paramsub$, and the complement of its space in $\params$
be $\paramsub'$, so $\paramsub \times \paramsub' = \params$. Elements in $\params'$ are
notated as $\parampt'$. Then Figure \ref{Fixfig} defines the \Fix transformation.

The \RNG and \CDF equivalences are by \refstmt{rnginvariance}. Because
this general case offers no relation between $\L(\parampt|\datapt)$ and
$\L'(\parampt'|\datapt, \parampt_f)$, we are unable to reuse $\Est$, although there
are many well-known special cases.

\subsection{Cross product} 
In an
example below, we will want a prior for the $\mu$ and $\sigma$ of a Normal distribution.
We could set the prior for $\mu$ as a Normal model and the prior for $\sigma$ as a
square-root-transformed $\chi^2$ model (see differentiable transformations,
below). These could be used sequentially, first setting $\sigma$ and then setting $\mu$,
or this transformation could be used to bind together the two models to produce a model
for $(\mu, \sigma)$.  

\newcommand{\Cross}{{\textsc{Cross}}\xspace}
\begin{figure}
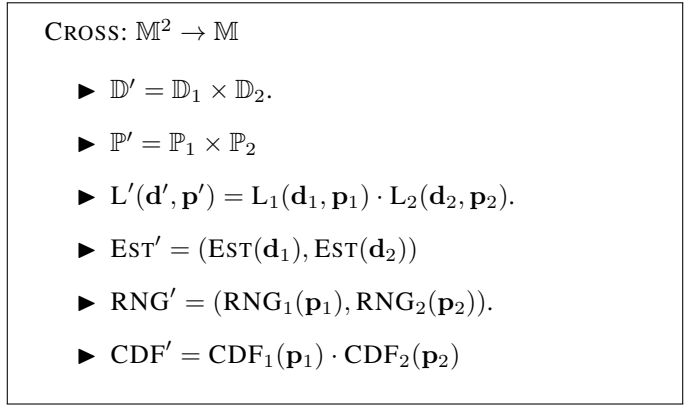

\begin{center}
\framebox[1.1\width]{
\begin{minipage}{8cm}
\vskip .1cm
\raggedright
\Cross: $ \models^2 \to \models $

\begin{itemize}\tighten
\def\labelitemi{$\blacktriangleright$}

\item $\datas'=\datas_1\times\datas_2$.

\item $\params' = \params_1 \times \params_2$
\item $\L'(\datapt', \parampt')=\L_1(\datapt_1, \parampt_1)\cdot \L_2(\datapt_2, \parampt_2)$.
\item $\Est'= \left(\Est(\datapt_1), \Est(\datapt_2)\right)$
\item $\RNG' = (\RNG_1(\parampt_1), \RNG_2(\parampt_2))$.
\item $\CDF' = \CDF_1 (\parampt_1)\cdot \CDF_2 (\parampt_2)$

\end{itemize}
\vskip .1cm
\end{minipage}
}
\end{center}

\caption{Definition of the \Cross mapping.}\label{Crossfig}
\end{figure}

The simplest case of binding together two models is where both have distinct
parameter spaces and data spaces. There is therefore effectively no interaction between
the models, and the independence assumption dictates that
$\L'(\datapt', \parampt')=\L_1(\datapt_1, \parampt_1)\cdot \L_2(\datapt_2, \parampt_2)$.
Figure \ref{Crossfig} specifies the rest of the \Cross transformation.

The \Cross morphism is associative, e.g., 
     $$\Cross(\mod{1}, \Cross(\mod{2}, \mod{3}))
     =\Cross(\Cross(\mod{1}, \mod{2}), \mod{3}).$$
Therefore, there is no ambiguity in writing 
$\Cross(\mod{1}, \mod{2}, \mod{3})$ to represent the combination of three models.

\newcommand{\Mix}{{\textsc{Mix}}\xspace}
\begin{figure}
\begin{center}
\framebox[1.1\width]{
\begin{minipage}{8cm}
\vskip .1cm
\raggedright
\Mix: $ (\models^2, w \in [0, 1]) \to \models $

\begin{itemize}\tighten
\def\labelitemi{$\blacktriangleright$}

\item $\datas' = \datas_1 = \datas_2$
\item $\params' = \params_1 \times \params_2\times [0,1] $
\item  $L_M(\datapt, \parampt_M) = w L_1(\datapt, \parampt_1) + (1-w) L_2(\datapt, \parampt_2).$

\item $\Est'$: Typically solved via EM algorithm \citep{dempster:em}.\footnote{The EM algorithm is a loop
consisting of the expected value function, \Fix,  and the \Est element of the parameter-fixed model.}
\item $\RNG'$: First, draw a random number $r\sim {\cal U}[0, 1]$. 
$$\RNG'(p_M) = \left\{\begin{array}{ll}
\RNG_1(\parampt_1) & r \leq w\\
\RNG_2(\parampt_2) & r > w
\end{array}\right.$$
\item $\CDF'(\parampt_w) = w \CDF_1(\datapt, \parampt_1) + (1-w)\CDF_2(\datapt, \parampt_2)$.

\end{itemize}
\vskip .1cm
\end{minipage}
}
\end{center}

\caption{Definition of the \Mix mapping.}\label{Mixfig}
\end{figure}

\subsection{Model mixing}
Consider two models, both over the same data space $\datas$ but distinct 
parameter spaces $\params_1$ and $\params_2$, and a weighting $w\in [0, 1]$, and let $\params_M$ be the Cartesian product
$\params_1 \times \params_2 \times [0, 1]$; in the other direction, one could decompose a point $\parampt_m$ into its three
parts, $\parampt_1$, $\parampt_2$, and $w$.  Figure \ref{Mixfig} displays the \Mix
transformation mapping from two models and a weight to a model representing the mixture.

As per examples to follow, 
the mixture transformation is especially useful in combination with the \Fix
transformation, for the purposes of fixing just the weights (fixing $w\equiv 0.5$ is a
popular choice), or fixing $\parampt_1$ and $\parampt_2$ and leaving the estimation only
to find the most likely $w$.

As with \Cross, one could recursively use the two-input \Mix transformation to
mix three or more models.

\subsection{Data Constraint} 
Let the constraint function $\Con_d:\datas \to \{0, 1\}$ be an indicator function that
is one when the input data is within some desired boundary and zero when it is not, and
let the constrained data space $\datas'$ be the set of data points where $\Con_d(\datapt)=1$.

\newcommand{\DataTrunc}{{\textsc{DataTrunc}}\xspace}
\begin{figure}
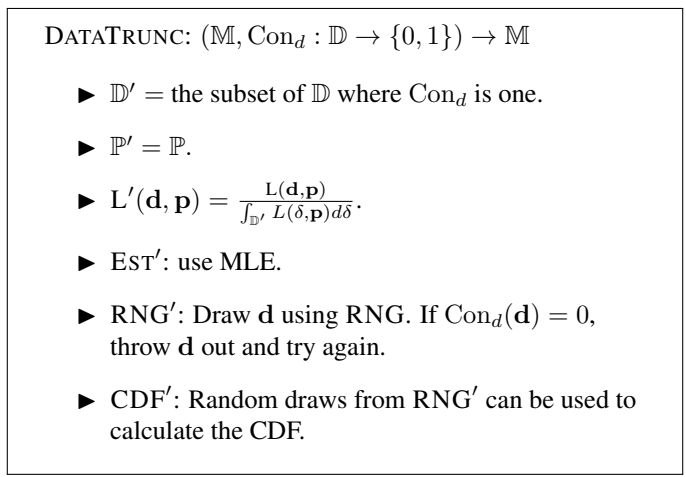

\begin{center}
\framebox[1.1\width]{
\begin{minipage}{8cm}
\vskip .1cm
\raggedright
\DataTrunc: $ (\models, \Con_d:\datas \to \{0, 1\})\to \models $

\begin{itemize}\tighten
\def\labelitemi{$\blacktriangleright$}

\item $\datas'=$ the subset of $\datas$ where $\Con_d$ is one.
\item $\params'=\params.$
\item $\L'(\datapt,\parampt)=\frac{\L(\datapt,\parampt)}{\int_{\datas'} L(\delta, \parampt) d\delta}$.
\item $\Est'$: use MLE.
\item $\RNG'$: Draw $\datapt$ using $\RNG$. If $\Con_d(\datapt)=0$, throw $\datapt$ out and try again.
\item $\CDF'$: Random draws from $\RNG'$ can be used to calculate the CDF.

\end{itemize}
\vskip .1cm
\end{minipage}
}
\end{center}

\caption{Definition of the \DataTrunc mapping.}\label{DataTruncfig}
\end{figure}

Figure \ref{DataTruncfig} describes a transformation, \DataTrunc, for the case where data outside the
constraint is suppressed. See \citet[\S 4.5]{cameron:count} for a discussion
of how maximum likelihood estimation of $\L_{\DataTrunc}$ recovers the parameters of
the unconstrained model,
and Listing \ref{roundtrip} for an example implementing and demonstrating the transformation.

\refstmt{rnginvariance} shows that within $\datas'$, $\RNG'$ and $\CDF'$ must be identical to
\RNG and \CDF; and it is easy to verify that $\RNG'$ is ML-consistent with $\L'$.

Because the scaling for \L depends upon \parampt, for an arbitrary unconstrained model $\mod{}$,
$\L_{\DataTrunc(\mod{})}(\parampt|\datapt)/\L_{\mod{}}(\parampt|\datapt)$
is not constant over all $\parampt$, so \refstmt{estinvariance} does not apply and the \Est routines will differ between pre- and
post-transformation models.

\paragraph{Example 6: A censored distribution with a point mass}

One could write the case of a truncated Normal distribution where any $\datapt < 0$ is
discarded using $M_{\rm trunc} = \DataTrunc_{x\geq 0}(M_{\cal N})$.

But consider a $\mod{\cal N}$ with all observations less than zero reported as exactly
zero.  One could use the models and transformation to this point to begin to express
this model, which is the combination of a PMF based on one data point
and a truncated Normal.

\begin{eqnarray*}
M_F &=& \Fix_{(w=1,\datapt=0)}(\mod{PMF})         \cr
M_T &=& \DataTrunc_{\datapt\geq 0}(M_{\cal N}).
\end{eqnarray*}

\newcommand{\Mixcdf}{{\textsc{Mixcdf}}\xspace}
\begin{figure}
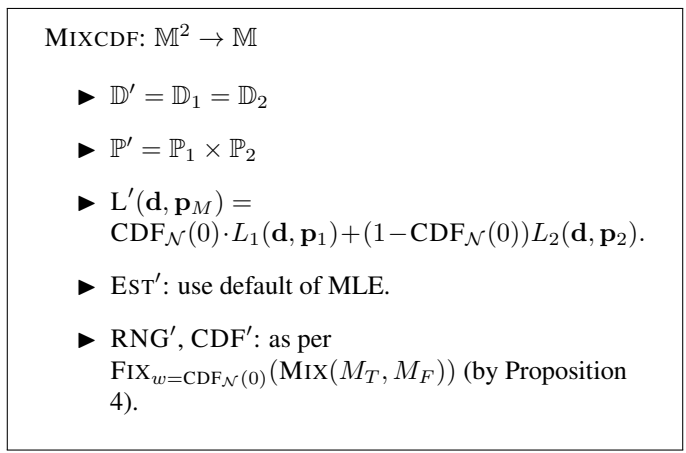

\begin{center}
\framebox[1.1\width]{
\begin{minipage}{8cm}
\vskip .1cm
\raggedright
\Mixcdf: $ \models^2 \to \models $

\begin{itemize}\tighten
\def\labelitemi{$\blacktriangleright$}

\item $\datas' = \datas_1 = \datas_2$
\item $\params' = \params_1 \times \params_2$
\item $\L'(\datapt, \parampt_M) = \CDF_{\cal N}(0)\cdot L_1(\datapt, \parampt_1) + (1-\CDF_{\cal N}(0)) L_2(\datapt, \parampt_2).$
\item $\Est'$: use default of MLE.
\item $\RNG'$, $\CDF'$: as per $\Fix_{w=\CDF_{\cal N}(0)}(\Mix(M_T, M_F))$ (by \refstmt{rnginvariance}).

\end{itemize}
\vskip .1cm
\end{minipage}
}
\end{center}

\caption{Definition of the \Mixcdf mapping.}\label{Mixcdffig}
\end{figure}

It is not quite correct to say that the final model is a mixture of $\mod{F}$ and $\mod{T}$,
because the mixing weight for $\Mix(\mod{F}, \mod{T})$ needs to be $\CDF(0, \mod{\cal N})$, which is
itself a function of $\mu$ and $\sigma$. We do not yet have a transformation that
sets a parameter at the value of a CDF, but it is trivial to write an {\it ad hoc}
transformation for the situation; see Figure \ref{Mixcdffig} for a description of the
\Mixcdf transformation.

Then the final model of a ${\cal N}(\mu, \sigma)$ with values less than zero replaced to
zero is $\mod{\rm Ncut} \equiv \textsc{Mixcdf}(\mod{T}, \mod{F})$.

\newcommand{\Jacobian}{{\textsc{Jacobian}}\xspace}
\begin{figure}
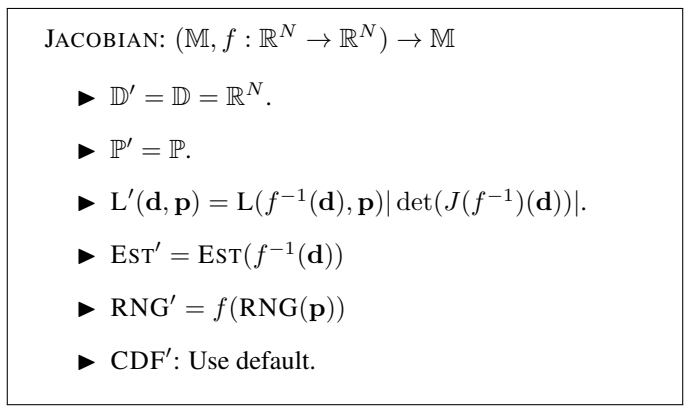

\begin{center}
\framebox[1.1\width]{
\begin{minipage}{8cm}
\vskip .1cm
\raggedright
\Jacobian: $ (\models, f:\Re^N \to \Re^N)\to \models $

\begin{itemize}\tighten
\def\labelitemi{$\blacktriangleright$}

\item $\datas' =\datas = \Re^N$.
\item $\params'=\params$.
\item $\L'(\datapt, \parampt)=\L(f^{-1}(\datapt), \parampt) |\det(J(f^{-1})(\datapt))|$.
\item $\Est' = \Est(f^{-1}(\datapt))$
\item $\RNG'=f(\RNG(\parampt))$
\item $\CDF'$: Use default.

\end{itemize}
\vskip .1cm
\end{minipage}
}
\end{center}

\caption{Definition of the \Jacobian mapping.}\label{Jacobianfig}
\end{figure}

\subsection{Differentiable transformation}
Consider a transformation
function $f:\Re^N \to \Re^N$, with inverse $f^{-1}:\Re^N\to\Re^N$, and inverse derivative matrix (the Jacobian)
$J(f^{-1})$. The likelihood in the mapped-to space is the likelihood in the
original space weighted by $|\det(J(f^{-1}))|$ 
See, for example, \citet{greene}. The transformation of likelihoods can naturally be
extended to a transformation of full models, as per Figure \ref{Jacobianfig}.

The set of invertible functions over a given space $\datas$ under the function
composition operator forms a group. Therefore, for a given base model $\mod{b}$,
the set of models $\Jacobian_{f}(\mod{b})$ over the same set of functions also form
a group under the model composition function defined by
$\Jacobian_{f}(\mod{b}) \circ \Jacobian_{g}(\mod{b}) \equiv \Jacobian_{f\circ g}(\mod{b})$.

Further, by the Radon–Nikodym theorem, for any two absolutely continuous models with
respect to a given measure, there exists a transformation between them. Thus, some
applications of other transformations in this paper are special cases of the \Jacobian transformation. 

\paragraph{Example 7: Volume}

 If the lengths of a set of cubes has a truncated-at-zero Normal distribution, $\DataTrunc_{\datapt>0}(\mod{\cal N})$, then the volumes
are described by $$\mod{\rm cube}= \Jacobian_{f(x)=x^3}(\DataTrunc_{\datapt>0}(\mod{\cal N})).$$

\newcommand{\Swap}{{\textsc{Swap}}\xspace}
\begin{figure}
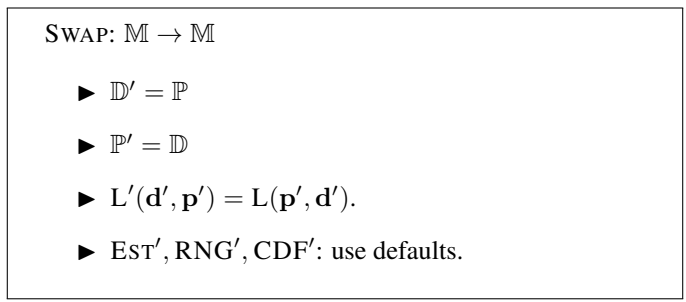

\begin{center}
\framebox[1.1\width]{
\begin{minipage}{8cm}
\vskip .1cm
\raggedright
\Swap: $ \models \to \models $

\begin{itemize}\tighten
\def\labelitemi{$\blacktriangleright$}

\item $\datas' = \params$
\item $\params' = \datas$
\item $\L'(\datapt',\parampt')=\L(\parampt', \datapt')$.
\item $\Est', \RNG', \CDF'$: use defaults.

\end{itemize}
\vskip .1cm
\end{minipage}
}
\end{center}

\caption{Definition of the \Swap mapping.}\label{Swapfig}
\end{figure}

\subsection{Swapping parameters and data}\label{swap}
Because data and parameters are not symmetric in the definition of a model, swapping them
produces a new model, as per the \Swap transformation defined in Figure
\ref{Swapfig}. Notably, the estimation process searches for the most likely parameters
given fixed data, so the post-swap model's estimation searches the orignal model's
data space given a fixed value in the original model's parameter space.

This will be used in Section \ref{mcmcrng} to implement random draws from a data space
via Markov Chain Monte Carlo and Section \ref{predsec} for finding the most likely
prediction from a model.

\subsection{Composition}

The transformations in this section consist of chaining together the output from one model (\mod{\rm from}) as
input to the next (\mod{\rm to}). Given that each model is associated with two spaces, there are four
possible means of chaining output from the first model to input to the second.

\begin{itemize}
\item If $\datas_{\rm from}$ is generated from the RNG of the first model,
linking 
$\datas_{\rm from} = \datas_{\rm to}$, allows us to evaluate the draws of one model using
the likelihood and estimation routines of another model.

\item In a hierarchical model, the parameter estimate from the child model(s) is used as
a data set for the parent model, in the form $\params_{\rm from} = \datas_{\rm to}$.

\item Matching $\datas_{\rm from} = \params_{\rm to}$ produces what is commonly referred to as
{\em Bayesian updating}.

\item The pairing $\params_{\rm from} = \params_{\rm to}$ is of limited utility.
\end{itemize}

\subsubsection{Data composition ($\datas_{\rm from} = \datas_{\rm to}$)}

The direction of $\datas_{\rm from} = \datas_{\rm to}$ is most commonly implemented by
using random draws from the first model as the input data set for the second. Let $Nseq
\in \Nat^N$
be an arbitrary sequence of natural numbers (in practice, draws from a PRNG).
Then $\RNG_{\rm to}(\datapt_{\rm to}) \in \datas_{\rm to}$, and because 
$\datas_{\rm from} = \datas_{\rm to}$, the draw can be used as an input to $\L_{\rm
from}$. Figure \ref{Dcomposefig} details the rest of this composition.

\newcommand{\Dcompose}{{\textsc{Dcompose}}\xspace}
\begin{figure}
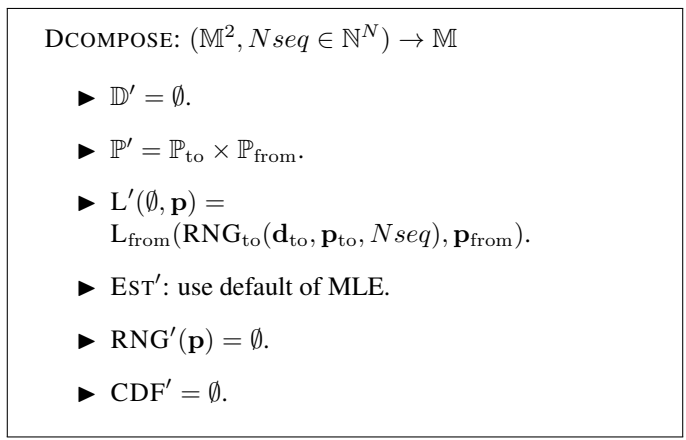

\begin{center}
\framebox[1.1\width]{
\begin{minipage}{8cm}
\vskip .1cm
\raggedright
\Dcompose: $ (\models^2, Nseq \in  \Nat^N)\to \models $

\begin{itemize}\tighten
\def\labelitemi{$\blacktriangleright$}

\item $\datas'=\emptyset$.
\item $\params' = \params_{\rm to}\times\params_{\rm from}$.
\item $\L'(\emptyset, \parampt)=\L_{\rm from}(\RNG_{\rm to}(\datapt_{\rm to}, \parampt_{\rm to}, Nseq), \parampt_{\rm from})$.

\item $\Est'$: use default of MLE.
\item  $\RNG'(\parampt) = \emptyset$.
\item $\CDF' = \emptyset$.

\end{itemize}
\vskip .1cm
\end{minipage}
}
\end{center}

\caption{Definition of the \Dcompose mapping.}\label{Dcomposefig}
\end{figure}

Attaching a fixed sequence of draws to the model retains the model structure: the
likelihood is deterministic and does not require additional PRNG inputs, and similarly for
the estimation routine. Each sequence $Nseq$ therefore defines a new model. In practice,
the class of models would likely be implemented on the fly, with random draws made as 
needed.  In such a case, \L and the MLE that depends on \L are stochastic. Some
ML search methods are better suited to stochastic search (e.g., simulated annealing, simplex
algorithms) than others (e.g., conjugate gradient searches).

\paragraph{Example 8: Exponentially-distributed random graphs}

Distributions of network link counts in human networks are known to take a few known
distributions, including the Exponential, Waring, Yule, or Zipf distributions.

So it is sensible to evaluate how well the simulated networks produced by
the network generation model above, $\mod{Sim}$, fit to an Exponential distribution, \mod{Exp}.

The network generation model produces data appropriate for 
composing with the Exponential distribution. The composed model, $$\mod{\rm
Ncomp}=\Dcompose_{Nseq}(\mod{Sim}, \mod{Exp}),$$ has $\datas=\emptyset$ and $\params =
\Re$, and can be used to evaluate the likelihood of the network link counts generated
by \mod{Sim} using the likelihood function of an Exponential distribution.

In the specification of Example 5, $\sigma$ was fixed at one, and so
$\params_{Sim}=\emptyset$; consider the
variant \mod{\sigma Sim} where $\sigma$ is not fixed, so $\params_{\sigma Sim}=\sigma$,
and \mod{Sim} = $\Fix_{\sigma=1}(\mod{\sigma Sim})$. What is the simulation that comes
closest to an Exponential model with $\lambda=1$? We could answer this question by
writing down the model $\mod{fix \lambda}=\Fix_{\lambda=1}(\mod{\sigma Sim}, \mod{Exp})$, and
using its estimate routine to find the optimal parameter $\sigma_{\rm opt}=\Est_{fix \lambda}$.
Evaluating $\Est_{fix \lambda}$ gives $\sigma\approx 0.51$.

\subsubsection{Parameter composition ($\datas_{\rm from} = \params_{\rm to}$)}\label{paramcompose}

This form of composition
is generally referred to by its immediate application: Bayesian updating.

For example, let  $\mod{likelihood} = \Fix_{\sigma=1}(\mod{\cal N})$ (so $\params_{\rm
likelihood}=\Re$,
representing $\mu$), and let a Normal distribution with known $\mu$ and $\sigma$
be the prior model (so $\datas_{\rm prior}=\Re$). Then the data space of the prior matches the
parameter space of the likelihood.  There are several ways to implement the elements
of this model, discussed below.

More generally, let $\mod{\rm prior}$ have $\rho \in \params_{\rm prior}$ and
$\parampt \in \datas_{\rm prior}$, and 
$\mod{\rm likelihood}$ have $\datapt \in \datas_{\rm likelihood}$
and
$\parampt \in \params_{\rm likelihood}$.  
The calculation of $\L_{\rm posterior}(\parampt|\rho, \datapt)$
can be broken down via Bayes's rule to be proportional to: 
\begin{equation}\label{br}
\L_{\rm prior}(\parampt|\rho)\cdot \L_{\rm likelihood}(\datapt|\parampt.
\rho),
\end{equation}

\newcommand{\DPcompose}{{\textsc{DPcompose}}\xspace}
\begin{figure}
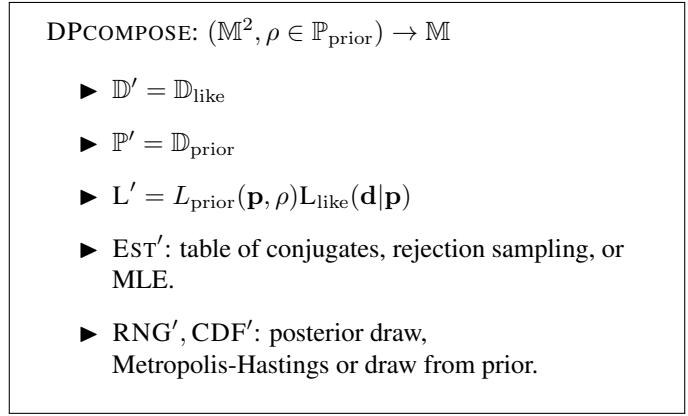

\begin{center}
\framebox[1.1\width]{
\begin{minipage}{8cm}
\vskip .1cm
\raggedright
\DPcompose: $ (\models^2, \rho \in  \params_{\rm prior})\to \models $

\begin{itemize}\tighten
\def\labelitemi{$\blacktriangleright$}

\item $\datas' = \datas_{\rm like}$
\item $\params' = \datas_{\rm prior}$
\item $\L' = L_{\rm prior}(\parampt, \rho)\L_{\rm like}(\datapt| \parampt) $

\item $\Est'$: table of conjugates, rejection sampling, or MLE.
\item $\RNG', \CDF'$: posterior draw, Metropolis-Hastings or draw from prior.

\end{itemize}
\vskip .1cm
\end{minipage}
}
\end{center}

\caption{Definition of the \DPcompose mapping.}\label{DPcomposefig}
\end{figure}

We take $\rho \in \params_{\rm prior}$ as a fixed-with-certainty prior belief regarding
the parameters of the prior model, $\datapt$ as being observed with certainty, and
use the fact that $\L$ need not integrate to one to specify the likelihood of the
posterior model. Figure \ref{DPcomposefig} details this data-parameter composition
transformation in full.

The default/model-specific set for estimating this transformation has three parts.

\paragraph{Conjugate priors} Some pairs of models are {\em conjugate
distributions}, which have known, closed-form posteriors. Tables of conjugate pairs are
readily available.\footnote{E.g.,
\url{http://en.wikipedia.org/wiki/Conjugate_prior#Table_of_conjugate_distributions}.}
In this case, the \Est, \RNG, \L, and \CDF elements of the prior-likelihood composition
can be replaced with the corresponding elements of the posterior model.

\paragraph{Likelihood prior}\label{mcmcrng} Metropolis-Hastings Markov chain Monte Carlo (MHMCMC)
can be used to draw from a distribution where the likelihood function can be easily
calculated to within some constant proportion, as in the case of Equation \ref{br}.
We could thus use it to make random draws from a prior-likelihood pair.
See \citet{metropolis} for details of the algorithm, which will be notated here as
a function $Metro: (\models, \models) \to \params$. Given this function, one can implement
Bayesian updating using the tools described to this point.

\begin{itemize}
\item Generate a model whose likelihood is the product of the input likelihoods via
\mod{c} = \DPcompose(\mod{prior}, \mod{likelihood}).
\item MHMCMC makes draws from the parameter space given fixed data, but we want to make draws
from data space given fixed parameters, so generate \mod{sc} =\Swap(\mod{c}).
\item MHMCMC requires a proposal distribution. A Normal distribution is a popular choice,
so let $\mod{propose}=\mod{\cal N}$. 
\item $Metro$ draws from the proposal and accepts or rejects the draw using a target
distribution, so we have enough to generate an \RNG for the model, drawing from the
data space of \mod{propose} and testing it against \mod{sc}.
\end{itemize}

\begin{figure}
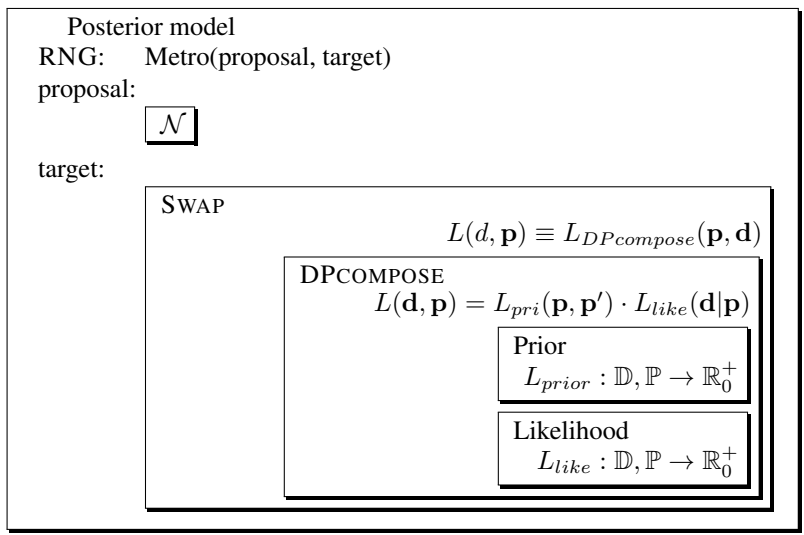


\cornersize*{.2mm}
\begin{center}
\shadowbox{
\shortstack{
\makebox[9cm][l]{Posterior model}\\
\hfill
\begin{tabular}{p{1cm}l}
\RNG:& Metro(proposal, target)\\
proposal:\\&  \shadowbox{
\shortstack{${\cal N}$}}\\
target:\\
&    \shadowbox{
\shortstack{  \makebox[8cm][l]{\Swap}\\
                \makebox[8cm][r]{$L(d, \parampt)\equiv L_{DPcompose}(\parampt, \datapt)$}\\
                \makebox[8cm][r]{\shadowbox{
\shortstack{
                        \makebox[6cm][l]{\DPcompose}\\
                        \makebox[6cm][r]{$L(\datapt, \parampt)= L_{pri}(\parampt, \parampt')\cdot L_{like}(\datapt|\parampt)$}\\
                        \makebox[6cm][r]{\shadowbox{
\shortstack{
                                    \makebox[3cm][l]{Prior}\\
                                    \makebox[3cm][r]{$L_{prior}: \datas, \params \to \Re^+_0$}
                                }}}\\
                        \makebox[6cm][r]{\shadowbox{
\shortstack{
                                    \makebox[3cm][l]{Likelihood}\\
                                    \makebox[3cm][r]{$L_{like}: \datas, \params \to \Re^+_0$}
                                }}}
                        }}}
            }}
\end{tabular}
}}
\end{center}

\caption{A posterior distribution augmented with an \RNG element based on
Metropolis-Hastings MCMC. Each box is a model; many transform the models
embedded within their box.}

\label{mcboxes}
\end{figure}

Figure \ref{mcboxes} is a diagram depicting the setup, a model where 
$$\RNG_{post} = Metro(\mod{\cal N}, \Swap(\DPcompose(\mod{prior}, \mod{like}))).$$

\shadowsize 1pt 

\paragraph{RNG prior} Consider a prior that has no explicit \L element, but does have an
\RNG element. We will build a PMF model for the posterior which, as above, has a
parameter set consisting of a list of pairs $(w_i, \parampt_i)$. For a fixed prior
parameter $\rho$, make a draw of $\parampt_i$ from the prior; by definition, this
has likelihood proportional to $\L_{\rm prior}(\parampt_i, \rho)$.  Let $\mod{\rm
fm}(\datapt, \parampt) =\Fix_{\parampt}(\mod{\rm like})$ and set the weight for this
datum as $w_i = \L_{\rm fm}(\datapt|\parampt_i)$. Then, as the count of elements
$\to\infty$, a draw from the PMF $(w_i, \parampt_i)$ is proportional to $\L'= \L_{\rm
fm}(\datapt|\parampt_i) d\L_{\rm prior}(\parampt_i, \rho)$.

MCMC methods are popular in part because the proposal distribution `walks around' in the
target distribution, spending more time in the regions with greater weight, so a
mismatched proposal (like a ${\cal N}(0, 1)$ proposal for a ${\cal N}(5, 1)$ target)
will adapt to become a well-matched proposal. Conversely the fixed prior model does
not move, and always makes more draws from the region(s) of the prior model with the
most weight. This can be inefficient if there is a large mismatch between the prior
and the likelihood. As the number of draws goes to infinity, drawing from the prior
directly and drawing from a proposal distribution would be equivalent, but for a
small number of draws, the MCMC chain generally gives smoother results. However, the
direct-draw method is feasible for RNG-based models for which MCMC is not, and is not
sequential. Typical computers in the present day can run several threads in parallel
(thousands at once in an increasing number of cases), and there may exist conditions where
a larger volume of parallel stationary draws could generate a more accurate posterior
than the adaptive sequential method.

\paragraph{Example 9: Bayesian updating with a simulation}

 Example 8 built \mod{Ncomp}
to describe a network simulation where the likelihood of a set of randomly
generated networks was described via an Exponential distribution.

We may have prior
beliefs that $\lambda \sim \Fix_{(\mu=m, \sigma=s)}(\mod{\cal N})$ for some known
parameters $(\mu=m, \sigma=s)$. Then we can generate a new model as

$$\mod{\rm post1} = \DPcompose(\Fix_{(\mu=m, \sigma=s)}(\mod{\cal N}), \mod{Ncomp}).$$

\begin{figure}
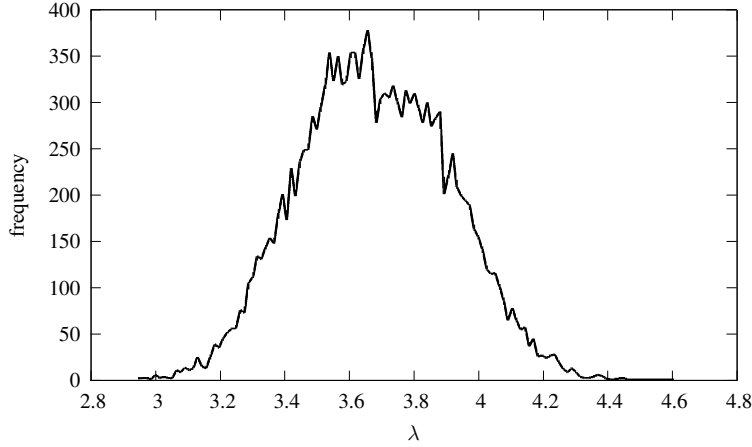

    \begin{center}
\scalebox{.8}{
\input lambda_post.tex
}
    \end{center}
\caption{A PMF representing the posterior distribution of $\lambda$ in the $\Dcompose_{Nseq}(\mod{Net}, \mod{exp})$ model.}\label{posteriorfig}
\end{figure}

Figure \ref{posteriorfig} shows the PMF reduction of this model.

Some authors draw a dichotomy between models that are on the table of conjugate
distributions, and can therefore be treated parametrically, and all other models whose
posterior must be approximated by a nonparametric distribution. However,
one can leave the model expressed using the prior's $\params$
and the likelihood's $\datas$. Here,

$$\mod{\rm post2} = \DPcompose(\mod{\cal N}, \mod{Ncomp}),$$

has $\params=\Re\times\Re^+_0$, representing $\mu$ and $\sigma$.  It is a parametric model
like any other, and its parameters can be estimated
using $\Est_{\rm post2}(\emptyset)$, and their confidence bounds can be calculated using
(stochastically appropriate) methods from Section \ref{testingsec}.

\subsubsection{Hierarchical modeling ($\params_{\rm child} = \datas_{\rm parent}$)}
\label{hi}

Consider a school, consisting of several classrooms. Each classroom provides a distinct
data set $d_1, \dots, d_N$, from which we can estimate the parameters of a model, such
as a set of regression coefficients or characteristics of a network model. Stacking
a model for each classroom produces $\mod{\rm classes}=$ $\Cross(M_{\rm class1},$
$\dots,$ $ M_{\rm classN})$, where $\Est_{\rm classes}(\datapt_1\times \dots \times
\datapt_N)$ is a stack of parameters
$(\Est(\datapt_1),$ $\dots,$ $\Est(\datapt_N))$ that can then be used as a data set for a
school-wide model.

\newcommand{\PDcompose}{{\textsc{PDcompose}}\xspace}
\begin{figure}
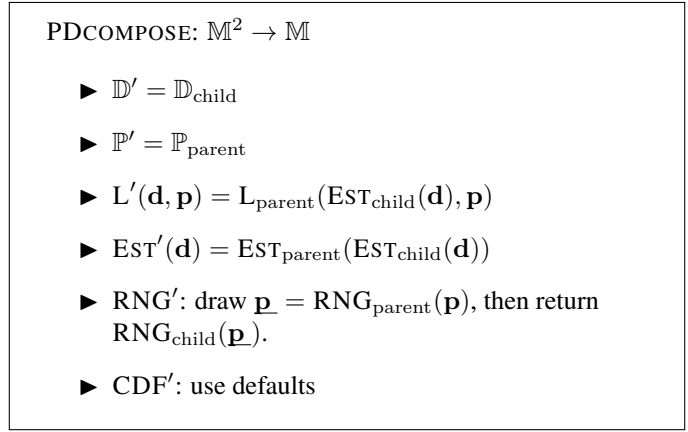

\begin{center}
\framebox[1.1\width]{
\begin{minipage}{8cm}
\vskip .1cm
\raggedright
\PDcompose: $ \models^2\to\models $

\begin{itemize}\tighten
\def\labelitemi{$\blacktriangleright$}

\item $\datas' = \datas_{\rm child}$
\item $\params' = \params_{\rm parent}$
\item $\LL'(\datapt, \parampt) =\LL_{\rm parent}(\Est_{\rm child}(\datapt), \parampt)$

\item $\Est'(\datapt) = \Est_{\rm parent}(\Est_{\rm child}(\datapt))$
\item $\RNG'$: draw $\underbar{\parampt} = \RNG_{\rm parent}(\parampt)$, then  return $\RNG_{\rm child}(\underbar{\parampt})$.
\item $\CDF'$: use defaults

\end{itemize}
\vskip .1cm
\end{minipage}
}
\end{center}

\caption{Definition of the \PDcompose mapping.}\label{PDcomposefig}
\end{figure}

It can be shown \citep{box:tiao} that in the case of linear models for both parent and
child, the model can be reduced to a single linear model. This is a special case of the
general form, where the parent and child models could take any arbitrary form, so long as
the parameter space of the child models match the data space of the parent model.
Figure \ref{PDcomposefig} describes this parameter-data composition transformation in detail.

\paragraph{Example 10: A narrative model}

The format of transforming simple models to produce more complex models suggests and
facilitates a style of narrative modeling, in which a real-world problem is broken
down into basic components and the manner by which the components are hypothesized to
combine is explicitly described via corresponding model transformations.

In this example, consider the arrival times of dinner party guests. One type of guest
tries to be on time, but may face delays, which occur at the rate of $\lambda$
per minute. The second type of guest tries to be fashionably late, ideally arriving 30
minutes late, but hitting that mark imprecisely. Nobody is early: those who miscalculate
and arrive in the neighborhood before the stated start time take care to delay enough to
arrive exactly on time.

The likelihood function for an Exponential distribution describes the percent of a group not yet exited from a
population given departures at a rate $\lambda$ as described above. The derivative of one
minus $\LL_{\rm exp}$ therefore describes the arrival rate at any given point in time, and
the reader can verify that this is 

$$M_A = \Jacobian_{1/\lambda}(\mod{\rm Exp}).$$

The second group is reasonably described
by a Normal distribution, with values less than zero fixed to exactly zero. This is
the \mod{\rm Ncut} model developed above. Assuming the two groups are evenly split,
the overall group is 

$$M_{\rm grp}=\Fix_{w=0.5}(\Mix(\Jacobian_{1/\lambda}(\mod{\rm Exp}), \mod{\rm Ncut})).$$ 

This model has $\datas=\Re^+_0$, representing minutes late; and
$\params=\Re^+_0\times\Re\times\Re^+$, for $(\lambda, \mu, \sigma)$.

We should put priors on these parameters. For $\lambda$, $\DataTrunc_{x>0}(\mod{\cal N})$
is a reasonable form for a prior; for $\mu$, we could use $\mod{\cal N}$; for $\sigma$,
we could use the square root of a $\chi^2$ model, $\Jacobian_{\sqrt{x}}(\mod{\chi^2})$ with parameters $\Sigma$. Then the prior is 
$$M_{\rm pri}=\Cross(
\DataTrunc_{x>0}(\mod{\cal N}), \mod{\cal N}, \Jacobian_{\sqrt{x}}(\mod{\rm InvWish}))$$

The overall model is then $M=\DPcompose(M_{\rm pri}, M_{\rm grp})$. It has parameters
$\params$ representing $(\mu_1, \sigma_1, \mu_2, \sigma_2, \Sigma)$, and $\datas=\Re^+_0$
representing minutes late.

There are a several things that one could do with $M$. One would be to fix the input
parameters to reasonable prior beliefs, use rejection sampling to reduce $M$ to a
nonparametric PMF, and compare the model to data using the data-space tests described in
Section \ref{applicationsec}. Or, one could estimate the five input parameters via $\Est(M)$
and run parameter-space tests (also discussed below) to give confidence bounds on the
prior parameters.

\section{Applications} \label{applicationsec}
This section lists a few existing methods that 
operate on generic black-box models. Any of the models described to this point could be
used for any of these applications.
This section assumes the reader is already familiar
with the methods, and will only briefly discuss the considerations needed to construct
algorithms around elements of \models, and their use for the broad range of models
that this paper contemplates.

As with many of the elements to this point, the goal is to establish an implementation for 
all models; closed-form implementations for certain models can be filled in as needed.

Before discussing a few more notable functions in detail, here are 
a few calculations that can work with input of a generic model (or models) in  a
straightforward manner:

\begin{itemize}
\item Entropy.
\item Kullback-Leibler divergence.
\item Cook's distance and other leave-$N$-out cross-validation techniques. 
\item Derivatives and second derivatives of differentiable parameters.
\end{itemize}

\subsection{Prediction}\label{predsec} The prediction problem is effectively a problem of partially
missing data. For example, given the independent variables of an OLS model, $\Xv$,
we may wish to predict the most likely values of the missing dependent variable $\Yv$.
We can generate a default algorithm for this problem using the tools above.

Given a model $M$ with known parameters $\parampt$, estimated using some reference or
training data, $\parampt=\Est(\datapt_{\rm ref})$, then $\Swap(M)$ generates a model where
the original $\parampt$ is treated as fixed data. $\Est_{\Swap(M)}(\parampt)$ produces the most
likely data for $M$.
Now consider a data set $\Datapt$, divided into known and unknown parts,
$\Datapt_{\rm known}$ and $\Datapt_{\rm miss}$. Then we can fix some parameters of $\Swap(M)$
at $\Datapt_{\rm known}$, leaving a model whose only unknown parameters are those
in $\Datapt_{\rm miss}$. Then the model is $\mod{pred}=\Fix_{\Datapt_{\rm known}}(\Swap(M))$,
and the maximum likelihood estimate of the data elements to be predicted is
$\Est_{pred}(\parampt)$.

\subsection{Testing} \label{testingsec}
Given that models in \models are an intermediary
between a data space and a parameter space, it makes sense to divide tests into those
regarding the data space and those regarding the parameter space.

\subsubsection{Parameter-space tests}
A test about a claim regarding the parameters has three steps:

\begin{enumerate}
\item Specify a statistic, such as a parameter estimate.
\item State the distribution of the statistic estimate as specified by the model.
\item Find the odds that the statistic lies within some given range of interest using the
CDF of the statistic as specified by the model.
\end{enumerate}

For example, after an OLS regression, the assumptions of the model indicate that the
parameter estimates are Multivariate Normal, and that fact is commonly used to state the
likelihood with which each parameter differs from zero.

We have a model's parameter estimates, from $\Est(\cdot)$, but also need the distribution
of the estimates. Setting aside those models (such as OLS) where it can be derived
via closed-form calculations, there are a few default methods available.

\paragraph{Covariance via bootstrap and jackknife} These methods take in a data set
and an estimation routine \citep{efron:gong,efron:tibshirani}. For iteration $i$ of
the algorithm, both methods generate an artificial data set (the bootstrap, via sampling
with replacement; the jackknife via a leave-$n$-out process), then run the estimation
routine to calculate $\hat\parampt_i$. Once several hundred or thousand such parameter
estimates are derived, the covariance matrix is calculated for the given parameters.

The final estimate of the statistic is the mean of the statistic calculated for each
data subset, so central limit theorems typically apply. These techniques
therefore apply to a wide range of models. However, they assume that draws from the data
are analogous to draws from the true population. 

\paragraph{Simple replication}
For the network model in Examples 5, 8, and 9, where
$\datas=\emptyset$, the parameter distribution can be produced by repeated draws
from the RNG. This is not bootstrapping, and does not rely on the core assumption
of bootstrapping: in a sense, the draws from the RNG could be thought of as the true
population asserted by the model.

\paragraph{Covariance via inverse Hessian} The Fisher Information matrix is the negation
of the inverse of
the expected Hessian matrix (the matrix of second derivatives of the log likelihood
function).\footnote{\citet{efron:hinkley} discuss the question of whether to use the
expected value of the Hessian or its value at the ML estimate only. They find that 
in typical cases, the value at the ML estimate is to be preferred. For some algorithms,
the Hessian is a computational side-effect of the maximum likelihood search, so its
computation is cheap.}
Especially for likelihood functions that are well-approximated by a Taylor expansion to
the second (squared) degree, Fisher Information can be a good estimate of the parameter
variance, and thus a good input to parametric hypothesis tests. Evaluating the quality of a second-degree Taylor
approximation in the neighborhood of an MLE can also be done using only the model
and a data set.

\subsubsection{Data-space tests}
For models where $\params$ is either trivial enough that it is $\emptyset$ or so complex
that the dimensionality of a given element $\parampt \in \params$ is unknown (i.e.,
for nonparametric models), it is difficult to use parameter-space tests. In this case,
we can develop metrics to describe the distance between models via computations over
the data space.

\label{metricsec}
One class of tests involve finding the distance from the fitted model to a target
model constructed by producing a PMF from a set of observed data, distinct from the
data used to estimate the model being tested. Given the two PMFs, there is then the
problem of summarizing their differences in a single statistic.

There are many options, and it it is beyond the scope of this paper to describe their
relative merits, but this segment discusses the problem of applying any of them to an
arbitrary element of $\models$.

Recall that the parameters of a PMF are a set of weights associated with points in the
data space: $\left((w_1, \datapt_1), (w_2, \datapt_2), \dots, (w_N, \datapt_N)\right)$.  
Given a second model that has different weights at the same data points---i.e., a model with parameters $\left((w_1', \datapt_1),
(w_2', \datapt_2), \dots, (w_N', \datapt_N)\right)$---one could calculate
the distance between the vector $(w_1, \dots, w_N)$ and $(w_1', \dots, w_N')$ by
a number of means:

\begin{itemize}
\item $k$-fold cross-validation tools typically report normalized Euclidian distance, aka root mean squared error, between the predicted and observed PMFs.
\item Kullback-Leibler divergence can be used to measure the information loss
from the actual to the predicted observations.
\item The Kolmogorov-Smirnov test reports a statistic based on the maximum distance between
the CDFs of the two distributions.
\end{itemize}

Given a PMF and an arbitrary distribution also defined over the same $\datas$, one
could construct a PMF via `binning'. The simplest method of constructing a PMF from an
arbitrary model $\mod{A}$ with parameters $\parampt$ is by setting $w_1=\LL_A(\datapt_1,
\parampt),$  $\dots,$ $w_N=\LL_A(\datapt_N, \parampt)$. Given a metric on $\datas$,
one could also define the set of points closest to $\datapt_j$ for each $j$, and
use $\CDF_A$ to calculate the total weight assigned to each set. Regardless of the
binning method chosen, the problem of comparing a discrete PMF to a continuous PMF is
now reduced to the problem of comparing two matched PMFs.

For two arbitrary models over the same $\datas$, we can more easily reduce the problem
to one of comparing matched PMFs, by drawing points from the data using either or
both $\RNG_1$ and $\RNG_2$, or sampling a grid of data points covering $\datas$
(provided the space is finite). Either method again reduces the problem to a comparison of
two matched PMFs.

\section{Conclusion} This paper presented a definition of a model as a collection
including a data space, a parameter space, $\Re^+_0$, and functions expressing the many
ways in which model users go between data and parameters.  Although this may seem
like a mere notational convenience, it provides sufficient structure to allow for the
construction of an extensive, internally consistent modeling language.

The examples in this paper demonstrate the egalitarian goals of the algebraic system,
as consistent functions and transformations are applied to models regardless of
whether they are Normal distributions, simulations, prior-likelihood combinations, or
nested combinations of all of these.  Given a set of qualitative results from
a simulation, it is natural to ask to which parameters the results are sensitive, and
this paper shows that there are sensible ways to use standard statistical techniques
for measuring variability given changes in parameters, and therefore the confidence
with which a parameter estimate can be made. Bayesian updating with a microsimulation
as a likelihood is not common in the literature, but is easy to operationalize in the
framework here, to the point of being almost obvious.

Writing down and applying the many model transformations presented in this paper was a
near-trivial matter thanks to the formal definition of a model.  The algebraic system
readily accommodates generic methods that are well-known in the literature, including
transformations via differentiable functions, mixture model estimation routines,
and so on. There are no technical limitations preventing the implementation of such
a system using virtually any modern computing platform.

Yet, in preparing this paper, I was unable to find a computing system (beside the
one demonstrated in the appendix) that offers a standard, genre-agnostic model form. Instead,
computing systems are typically built around structures that are most convenient to
immediate problems.  This paper is a demonstration of how much can be built from a
consistent model object, and an invitation to authors to consider making greater use
of model objects such as the one described here as they develop new platforms or build new
models using existing platforms.

\section*{Appendix}
This appendix presents some technical notes on the implementation of a model object, and
a number of examples of model creation and manipulation. The examples are in standard
C using the open-source Apophenia library of functions for scientific computing, which may create
difficulties for those readers who have limited proficiency in C or no interest in using
Apophenia.\footnote{{\em Standard C} means code conforming to the ISO/IEC 9899:2011
standard.} The hope is that such readers will still be able to follow how models are
created, transformed, and used, even if the details of syntax are unfamiliar.
Readers who would like more on the details of syntax are referred to the Apophenia
documentation, at \url{http://apophenia.info}.

Although the examples are in C, {\bf these routines could be written in any Turing-complete
programming language}. Readers proficient with other platforms are encouraged to consider
how these programs could be written using their preferred platforms.

Despite the difficulties, there are benefits to showing complete code examples.
First, many of these examples produce uninteresting output; it is the process by
which the output was produced that is of interest.  Second, this code compiles and
runs, demonstrating that the concepts in this paper can be beneficially applied to
real-world problems. Third, pseudocode and mathematical reductions may have hidden
errors or omissions; using tested code, we can be much more confident that all relevant
considerations have been addressed.

\paragraph{Notes on implementing a model object} The algebraic system in this paper
focuses on a single object class, the model. Before presenting the examples, this
segment first presents a few notes on the details of how the class of {\tt apop\_model} 
objects is implemented.

The object is a structure collecting several elements:

\begin{itemize}
\item Data
\item Parameters
\item Several functions, including those in Definition \ref{modeldef}, but also 
a constraint function for doing constrained optimization and some
functions to handle logistics. There are both \verb@p@ and \verb@log_likelihood@ functions, which are largely redundant, but allow authors to use whatever is more common in their genre of modeling.
\item A hook for groups of settings, such as a group describing how a maximum likelihood
search should be run, or a collection of the items requisite for running a simulation.
\item A flag for marking processing errors.
\end{itemize}

Apophenia's models are a C {\tt struct} holding
these elements. For example, the Normal distribution model is encapsulated in the {\tt
apop\_normal} structure, which largely depends on the GNU Scientific Library (GSL)
for the computational work \citep{gough:gsl}. The implementation of the Multivariate
Normal is very similar, and largely uses the GSL as well. Apophenia implements the
univariate and multivariate cases
separately, though one could in fact implement only the Multivariate Normal, because
it reduces to a univariate Normal given one-dimensional input.

The {\tt estimate} routine has the form $\Est:(\datas, \models) \to \models$, which
differs from the estimation function described above.  The parameters of the input
model are {\tt NULL}; the output model is a copy of the input model with the parameters
set.

Apophenia implements default routines for estimation, random draw, and other methods
via dispatch functions. For example, here is a short program to
read in a data set (in plain text format, whose first column is the dependent variable
and subsequent columns are numeric independent variables), estimate an OLS model,
and display the estimated model to the screen.\\
{\tt \#include $<$apop.h$>$\\int main()\{\\\ttab    apop\_data *d = apop\_text\_to\_data("dataset.csv");\\\ttab    apop\_model *estimated = apop\_estimate(d, apop\_ols);\\\ttab    apop\_model\_print(estimated, NULL);\\\}}

The {\tt apop\_estimate} function checks the input struct, {\tt apop\_ols} for a non-null
{\tt estimate} function. If one is present, then the data is sent to that function. If
the input struct has a null {\tt estimate} function, then the data and model are sent
to the default routine: {\tt apop\_maximum\_likelihood}.\footnote{
Because maximum likelihood search is the the default for estimation,
Apophenia includes several optimization methods borrowed from the GSL, including a few
types of conjugate gradient methods, Newton's method, the Nelder-Mead simplex algorithm,
and simulated annealing. The model object may have an associated score function (the
dlog likelihood), which is used by some of the search methods.  Given a model {\tt m}
with no closed-form score, it is sometimes better to use a non-derivative method,
and sometimes better to use {\tt apop\_numerical\_gradient(m)} to approximate the
score of {\tt m} via delta method.

Some large-dimension searches work best via a per-dimension search: fix all parameters
at their expected value given the input model; do a search for the optimum of the
first parameter; fix that parameter at its optimum; search for the optimum of the
second parameter; repeat until the end of the parameter list; repeat the loop until
the change in objective function over a search is less than a user-specified tolerance.
Implementing this simply requires repeated application of the \Fix transformation.}

\citet{r:scoping} produce models like {\tt estimated} via {\em closures}, which are
functions bound together with an environment holding variables used in the function.
Similarly, the {\tt apop\_model} is a struct holding both the functions in Definition
\ref{modeldef} and the now-estimated parameters (and potentially other model-specific
settings, discussed below).

As an aside, all new functions in any language and for
any purpose need to be tested, which can be difficult for numeric algorithms beyond
nontrivial data. Having a default method and a model-specific method that theoretically
achieve the same result provides a sensible testing procedure: test whether the default
method and model-specific methods produce results within acceptable tolerances of
each other.

Some models and methods require variables and settings not listed in the core model
struct. The \verb@apop_model@ struct can therefore hold a list of {\em settings
groups}.  Functions are provided for adding, modifying, and removing settings groups
to a model.  For example, a histogram model might have a settings group
specifying bin locations.

To give a more involved example, there are effectively two
use-cases for an MCMC chain. One is to generate a PMF by making a large number of
draws from the chain and recording them as a batch. The other is to make individual draws
from the chain, in the style of a typical random number generation function. A settings
group attached to the model from which draws are made holds the user-specified burnin
period, the chain up to that point, and its final state. For use as a model, the entire
run to that point is available; for use as an RNG, the state is available for making the
next draw from the burnt-in chain.

Many of the transformations described in the body of this paper output a model specially
written for the transformation with a settings group pointing to the base model.
For example, the input to {\tt apop\_model\_fix\_params} is a model estimated using a parameter set
with some numeric values and some NaN values, where NaN is a not-a-number semaphore
as specified in the IEEE 754 standard. Parameters with not-NaN values are fixed at
the given values; parameters set to NaN are left free.  
The output from the transformation is a model with a settings group pointing to the
original model. The output model has \Est, \LL, \RNG, \dots functions that do the
requisite transformation on the inputs, call the base model, and return the (possibly
transformed) result from the base model's \Est, \LL, \RNG, \dots.

The list of methods embedded in the model struct is deliberately restricted. Other things
one might do with a model, including prediction, specifying submodels for estimated parameters,
calculating the score, and printing to the screen or a file, are implemented via a 
{\em virtual table} (vtable) mapping keys to functions. The keys are built by
combining relevant elements of the model. E.g., for a Bayesian updating routine,
the relevant elements are the likelihood functions of the prior and likelihood. If those
functions appear on a table of conjugates, the vtable finds a routine that produces the
appropriate output, and if the likelihoods are not found in the vtable, the default of
MCMC or drawing from the prior is used (depending on which elements are non-null in the
prior).

\paragraph{Example 11: An RNG--estimation round trip}

This first set of examples includes a truncation transformation function for models where
$\datas=\Re$, and a set of uses.  The uses fulfill the promise of egalitarian treatment
of models both before and after truncation.

Generally, the code examples here can be read from the bottom up, as the last function
in the example will call functions earlier in the listing, which may call functions
earlier still.  The last function of Listing \ref{trunc},  \verb@truncate_model@,
takes in a model and returns a model truncated at a given cutoff.

The remainder of the listing defines a model to be returned. The \verb@prep@ function does
Apophenia-specific work, setting the output model's parameter sizes, and storing the
original, untruncated model at the output model's \verb@more@ pointer. The RNG
for the truncated model makes draws from the model stored at \verb@more@ until one of
the draws is greater than or equal to the cutoff.

The example implements the \DataTrunc transformation for the special case where
$\datas=\Re$ and the cutoff is of the form $f(\datapt)=1$ iff $x \geq k$ for some cutoff $k$.
Note that the truncated model produced by the transformation is lacking an estimation
routine, so that must be filled in via defaults.

\longcodefig{149-truncate}{A truncation transformation.}{trunc}

Listing \ref{roundtrip} demonstrates a
test of a model, in which a few thousand draws are made using the model's RNG, and then
that drawn data set is used as input to the model's estimation routine. If the parameters
estimated match the parameters used at the beginning of the process, our confidence that
the RNG and estimation are consistent rises. The notable point about this code is that
\verb@main@ calls this round-trip function in the same manner with four models: a Normal distribution, a truncated Normal, a Beta
distribution, and a truncated Beta distribution. 

The output of the program will not be printed here because, as with most of the examples
in this appendix, the actual output is as expected or trivial. In this case, it prints
the estimated $\mu$  and $\sigma$ for the Normal and truncated Normal---(1.000660,
0.999806) and (0.998964, 0.997244), respectively---and $\alpha$ and $\beta$ estimates
for the Beta and truncated Beta---(0.700536, 1.706045) and (0.716461, 1.717567),
respectively.

\longcodefig{149-roundtrip}{A series of round trips.}{roundtrip}

\paragraph{Example 12: Estimating the parameters of a  prior+likelihood model}

In this example, Nature generated data using a mixture of
three Poisson distributions, with $\lambda=2.8$, $2.0$, and $1.3$. Not knowing the
true distribution, the analyst wrote down a model with a truncated Normal(2,
1) prior describing the parameter of a Poisson likelihood model (\mod{\cal P}).  That model produces a
posterior distribution over $\lambda$. The analyst would like to present an approximation
to the posterior in a simpler form, and so
finds the parameters $\mu$
and $\sigma$ of the Normal distribution that is closest to that posterior, given the
data.

The storyline can be expressed as single function:
\begin{align*}
&&\Est(
        \RNG\left(\Mix\left(\Fix_{2.8}(\mod{\cal P}),
        \Fix_{2.0}(\mod{\cal P}),
        \Fix_{1.3}(\mod{\cal P})\right)\right),\hfill\\
 &&       \DPcompose\left(\DataTrunc_{d>0}(\Fix_{\mu=2, \sigma=1}(\mod{\cal N}),
                        \mod{\cal P}\right),
\mod{\cal N}).
\end{align*}
Listing \ref{updatefn} is is the transcription of this function.

\longcodefig{150-update}{A Normal approximation to an updated model, in a functional style.}{updatefn}

This is the `functional' style of expression, where the program execution is 
the evaluation of a single complex function. The one difference from textbook functional
code is that the use of the \RNG function makes the program stochastic: setting
\verb@apop_opts.rng_seed@ to an arbitrary integer would cause a small change in the
final evaluation.

\paragraph{Example 13: A demand-side ABM}

 Listing \ref{demandsim} is a simple model of agents
deciding the quantity of goods to buy given prices, preferences, and a budget. It
is effectively an RNG, but Listing \ref{demand} does the same round-trip as earlier
examples, generating a data set using the model's RNG, and then estimating the optimal
parameters of the model given that data set.

The model:

\begin{items}
\item There are 1,000 agents.
\item Each agent $i$ has a budget allocation $b_i$ and preference coefficient $\alpha_i$, each drawn from independent
Normals. Fix $\sigma=1$ for both distributions, leaving us with
two parameters from this part of the model: $\mu_b$ and $\mu_\alpha$.

\item Two goods are available for purchase; the first has price $p$ and the second price one
(without loss of generality from a setup with two prices $p_1$ and $p_2$).

\item Each agent's utility from purchasing the good is $U(q_1, q_2, \alpha) = q_1^{\alpha} + q_2$. Agents are utility maximizing.

\item  Agents maximize subject to the budget constraint that $pq_1+q_2 \leq b_i$.

\item We observe the mean consumption $Q_1$ and $Q_2$ (total consumption divided by the number of agents).

\end{items}

For this simple case where there are no interactions among agents, the consumption problem can be easily solved.
Given the problem
\begin{align*}
    \max u &= q_1^\alpha + q_2\\
    \ni b &= p_1q_1 + q_2,
\end{align*}
the optimum, where $\partial u/\partial q_1 = 0$ and the budget constraint is
        satisfied, is 
\begin{align*}
q_1 &= (p_1/\alpha)^{1/(1-\alpha)}\\
q_2 &= \max(b - p_1q_1, 0).
\end{align*}

The core of the model, in Listing \ref{demandsim}, is a single draw of a population
and its decisions, in the {\tt draw} function. The first several lines shunt parameters,
then the agents are drawn, then the next loop does the optimization for each agent.
This form is not very streamlined, but is easily extensible; for example, we might
want to have a step where agents interact between the loop generating the population
and the loop calculating their consumption decisions.

\longcodefig{154a-demand_curve}{The core of a demand-side model.}{demandsim}

Listing \ref{demand} wraps the simulation into a model.  
Looking at the {\tt p} function, we see that a single likelihood for a given
data/parameter set is found by making 500 draws, then smoothing the PMF using a kernel
density estimate in which a Multivariate Normal is placed over every drawn point. The
bulk of the code in this segment is spent setting up the Multivariate Normal and KDE.

The {\tt main} function executes a simple model application similar to the
round-trips as in Listing \ref{roundtrip}, generating a small data set, then
estimating the optimal parameters given that data set.  Setting the \verb@dim_cycle@
element in the MLE settings group tells the optimizer to use the EM-style strategy of
dimension-by-dimension optimization.

\longcodefig{154-demand_curve}{Setting up and using the \RNG, \LL, and \Est elements of the demand-side ABM.}{demand}

\paragraph{Example 14: A search model}

Listing \ref{seek} is a spatial search model:

\begin{items}
\item An equal number of agents of types A and B are randomly placed on a grid. They seek to
pair with an agent of opposite type.
\item Agents in the middle of the grid have eight neighboring squares; agents on the edge or a
corner have fewer because they are unable to go off the grid.
\item Until all agents are paired up:
    \begin{items}
    \item If an agent is adjacent to another agent of the opposite type, they pair up,
    and are taken out of the model. We record how long it took for them to find each other.
    \item Remaining agents take a single step to a random unoccupied neighboring square.
    \end{items}
\item The model output is the list of pairing times.

\end{items}

We can expect that initially, when there are many agents on the grid, that
some agents will find each other quickly. Eventually, there will be only one A
agent and one B agent, and they will wander for a long time before pairing up.

The ratio of agent count to grid size matters: 90 agents on a
$10\times 10$ grid is a very different search from 90 agents on a $1,000 \times 1,000$
grid. We will put a prior on these below.

Much of the code is simple and mechanical; Listing \ref{seekfns} presents some macros and
functions for use by agents searching for a mate or stepping to another cell in the grid.
This listing is included for completeness, but is especially C-specific and can be skipped.
The last line of Listing \ref{seek} wraps the simulation as a model $\in \models$, with
$\params=\emptyset$ and $\datas$ representing the pairing times of each agent.

\begin{items}
\item There is a random number generator associated with every agent. RNGs tend to not be
parallelizable, but if every agent has its own, then one could still thread the agents'
activities.

\item The agents live in two structures, set up in \verb@run_sim@. One is a non-changing array
of agents. The second is a grid of pointers, with either {\tt NULL}
(vacant) or a pointer to one of the agents in the unchanging list of agents. We can easily
loop through the agents via the array, and check positions and their surroundings with the
grid. When agents are paired up, mark {\tt done=true} in the agents' records, leave them in
the array, and erase their presence on the grid.
\end{items}

\longcodefig{156-search_fns}{Mechanical functions for use by the search model in Listing \ref{seek} as {\tt search\_fns.c}. Included for completeness.}{seekfns}

\longcodefig{156-search}{Agents searching for each other on a grid.}{seek}

One run produces a list of agent exit times. The rate at which agents exit
changes with time. This is also the story of a Weibull distribution, so it would be interesting
to see how well such a distribution fits. A Weibull has two parameters, $k$ and
$\lambda$.  If $k=1$, then $\lambda$ is the mean time to exit. If $k<1$, then the time
to exit is slower for those that are still present later in the game: some (probably the
more centrally-located) get picked up quickly, and the hard-to-match take several
$\lambda$s' time to find a match.  So for this simulation, we expect that $\lambda$
grows as the grid gets more sparse, and $k$ should be noticeably less than one.

Listing \ref{weibull} specifies a Weibull model. As with most models that
assume iid data, the log likelihood of one element can be written down, and then the total
log likelihood is the sum of the map of that function onto every element. The
\verb@apop_model@ struct includes 
a constraint to keep the MLE search routine from setting $\lambda$ or $k$ to zero.

\longcodefig{156-weibull}{A Weibull model.}{weibull}

In listing \ref{find}, the \verb@one_run@ function sets the settings for the simulation,
makes a few thousand draws, and estimates the parameters of a Weibull using those draws,
thus giving one point estimate of the parameters.

\longcodefig{156-find}{Fitting a Weibull distribution to data from the
search model}{find}

\begin{figure}
\begin{center}
\scalebox{.4}{
    \batchmode
\includegraphics{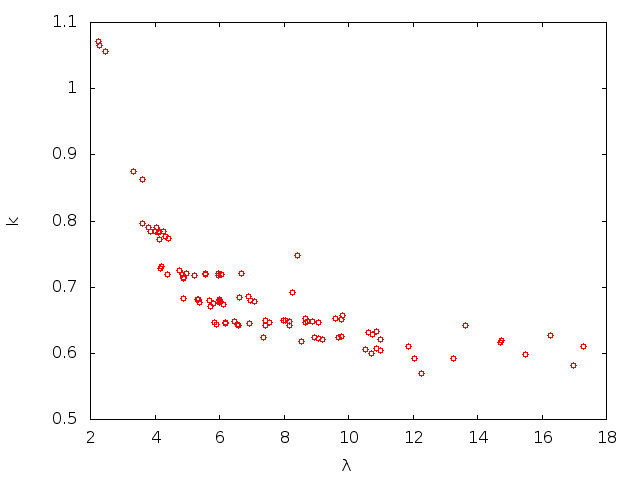}
\errorstopmode
}
\end{center}
\caption{The Weibull parameters given simulation inputs with independent Normal distributions.}\label{lambdak}
\end{figure}

The \verb@fuzzed@ function puts a prior on the grid size and population settings.
The output is a PMF of 100 Weibull parameters.  Figure \ref{lambdak} is a plot of
draws from the posterior distribution of $(\lambda, k)$.

 \setstretch{1.4}
\bibliographystyle{plainnat}
\bibliography{modeling}
\end{document}